\documentclass[twocolumn]{aastex62}

\usepackage{booktabs}
\usepackage{longtable}
\usepackage{appendix}
\usepackage{amsmath}
\usepackage{amssymb}
\usepackage{graphicx}
\graphicspath{{figures/}}

\shorttitle{Mintz et al.}
\shortauthors{Mintz}

\def\gamamerianmatches{65,542}
\def\sdssmerianmatches{134,556}
\def\sdssandgamamerianmatches{9,209}
\def\merianmerianmatches{1,153}

\def\merianspecsample{2,089}

\def\specsamplesize{180,485}
\def\specsamplesizeinband{9,774}
\def\specsamplesizeinbandlowmass{4,646}
\def\specsamplesizeinbandlowmassallphot{2,311}

\def\failedfit{204}

\def\samplesizegoodfit{1,754}

\def\morphremove{277}
\def\artifacts{200}
\def\staroverlap{27}
\def\postmorphsample{1,250}

\begin{document}

\title{A Nonparametric Morphological Analysis of H$\alpha$ Emission in Bright Dwarfs Using the Merian Survey}

\author[0000-0002-9816-9300]{Abby Mintz}
\affil{Department of Astrophysical Sciences, Princeton University, 4 Ivy Lane, Princeton, NJ 08544, USA}

\author[0000-0002-5612-3427]{Jenny E. Greene}
\affil{Department of Astrophysical Sciences, Princeton University, 4 Ivy Lane, Princeton, NJ 08544, USA}

\author[0000-0002-0332-177X]{Erin Kado-Fong}
\affil{Physics Department, Yale Center for Astronomy \& Astrophysics, PO Box 208120, New Haven, CT 06520, USA}

\author[0000-0002-1841-2252]{Shany Danieli}
\affil{Department of Astrophysical Sciences, Princeton University, 4 Ivy Lane, Princeton, NJ 08544, USA}
\affil{School of Physics and Astronomy, Tel Aviv University, Tel Aviv 69978, Israel}

\author[0000-0001-9592-4190]{Jiaxuan Li}
\affil{Department of Astrophysical Sciences, Princeton University, 4 Ivy Lane, Princeton, NJ 08544, USA}

\author[0000-0001-7729-6629]{Yifei Luo}
\affil{Department of Astronomy and Astrophysics, University of California, Santa Cruz, 1156 High Street, Santa Cruz, CA 95064, USA}

\author{Alexie Leauthaud}
\affil{Department of Astronomy and Astrophysics, University of California, Santa Cruz, 1156 High Street, Santa Cruz, CA 95064, USA}

\author{Vivienne Baldassare}
\affil{Department of Physics and Astronomy, Washington State University, Pullman, WA 99163, USA}

\author{Song Huang}
\affil{Department of Astronomy, Tsinghua University, Beijing 100084, China}

\author[0000-0002-8040-6785]{Annika H. G. Peter}
\affil{Department of Physics, The Ohio State University, Columbus, OH 43210, USA}
\affil{Department of Astronomy, The Ohio State University, Columbus, OH 43210, USA}
\affil{Center for Cosmology and Astro-Particle Physics, The Ohio State University, Columbus, OH 43210, USA}

\author{Joy Bhattacharyya}
\affil{Department of Astronomy, The Ohio State University, Columbus, OH 43210, USA}
\affil{Center for Cosmology and Astro-Particle Physics, The Ohio State University, Columbus, OH 43210, USA}

\author{Mingyu Li}
\affil{Department of Astronomy, Tsinghua University, Beijing 100084, China}

\author{Yue Pan}
\affil{Department of Astrophysical Sciences, Princeton University, 4 Ivy Lane, Princeton, NJ 08544, USA}



\begin{abstract}
Using medium-band imaging from the newly released Merian Survey, we conduct a nonparametric morphological analysis of H$\alpha$ emission maps and stellar continua for a sample of galaxies with $8\lesssim\log (M_\star/M_\odot) < 10.3$ at $0.064<z<0.1$. We present a novel method for estimating the stellar continuum emission through the Merian Survey's N708 medium-band filter, which we use to measure H$\alpha$ emission and produce H$\alpha$ maps for our sample of galaxies with seven-band Merian photometry and available spectroscopy. We measure nonparametric morphological statistics for the H$\alpha$ and stellar continuum images, explore how the morphology of the H$\alpha$ differs from the continuum, and investigate how the parameters evolve with the galaxies' physical properties. In agreement with previous results for more massive galaxies, we find that the asymmetry of the stellar continuum increases with specific star formation rate (SSFR) and we extend the trend to lower masses, also showing that it holds for the asymmetry of the H$\alpha$ emission. We find that the lowest-mass galaxies with the highest SSFR have H$\alpha$ emission that is consistently heterogeneous and compact, while the less active galaxies in this mass range have H$\alpha$ emission that appears diffuse. At higher masses, our data do not span a sufficient range in SSFR to evaluate whether similar trends apply. We conclude that high SSFRs in low-mass galaxies likely result from dynamical instabilities that compress a galaxy's molecular gas to a dense region near the center.  
\end{abstract}



\section{Introduction} \label{sec:intro}

In recent years, there has been a revolution in the study of spatially resolved star formation in Milky Way--analog galaxies. Surveys including the Sydney-AAO Multi-object Integral Galaxy survey \citep[SAMI;][]{Croom2012}, the Calar Alto Legacy Integral Field spectroscopy Area survey \citep[CALIFA;][]{Sanchez2012}, Mapping Nearby Galaxies at APO \citep[MaNGA;][]{Bundy2015}, and Physics at High Angular resolution in Nearby GalaxieS \citep[PHANGS;][]{Leroy2021, Emsellem2022, Lee2022, Lee2023} have amassed spatially resolved data over the full electromagnetic spectrum, simultaneously tracing the star formation, molecular gas, dust extinction, star cluster formation, and gas-phase metallicity, among numerous other critical physical properties of nearby galaxies. These advances have led to substantial progress in our theories of galaxy-wide star formation. Still, due to observational limitations, it has not been possible to replicate these measurements to the same extent for dwarf galaxies. 

Dwarf galaxies represent a particularly exciting demographic for testing our current theories of star formation; they are gas-rich \citep{Bradford2015}, but yet inefficient at forming stars \citep{Leroy2008}, bursty \citep{Weisz2012, Guo2016, Sparre2017, Emami2019, Atek2022,Pan2023}, and significantly more impacted by feedback and environment than their more massive counterparts \citep{MacLow1999, Christensen2016, ElBadry2016}. 

Surveys of H$\alpha$ and UV emission like the 11 Mpc H$\alpha$ UV Galaxy Survey \citep[11HUGS;][]{Kennicutt2008, Lee2011} and the Legacy Extragalactic UV Survey \citep[LEGUS;][]{Calzetti2015} have helped to thoroughly characterize star formation in local dwarfs \citep[e.g.,,][]{Lee2007, Lee2009a, Lee2009b,Lee2016, Cignoni2018, Hunter2018a, Hunter2018b,Cook2019} but have had limited statistics owing to their focus on nearby galaxies. A number of other surveys have mapped H$\alpha$ emission using narrowband or medium-band filters \citep[e.g.,][]{GildePaz2003, Meurer2006, Gavazzi2012, Boselli2015, Gavazzi2018,Griffiths2021, HermosaMunoz2022, Terao2022, Sieben2023, Chen2023}, but most have focused on higher-mass galaxies, and they are nearly always confined to the Local Volume ($D<10\,\mathrm{Mpc}$). To date, no survey has extensively mapped resolved H$\alpha$ emission in dwarfs beyond 10 Mpc.

The newly conducted Merian Survey \citep[Danieli et al. 2024, in review]{Luo2023} provides a unique opportunity to study these intriguing galaxies and probe fundamental questions about galaxy-scale star formation using two-dimensional distributions of star formation in low-mass galaxies outside the Local Volume. The primary goals of the survey are to provide accurately measured photometric redshifts for $\sim$85,000 dwarf galaxies with $8<\log (M_\star/M_\odot)<9$ and to measure their weak gravitational lensing signal. The survey will map the footprint of the Hyper Suprime-Cam Subaru Strategic Program \citep[HSC-SSP;][]{Aihara2018} using two optical medium-band filters designed to capture [\ion{O}{3}] and H$\alpha$ emission for galaxies at $0.06\lesssim z \lesssim 0.1$. In addition to reliable distance measurements, the medium-band filters also provide a wealth of information about spatially resolved star formation in the imaged galaxies within the redshift range. 

\begin{figure*}[t]
\begin{center}
\includegraphics[width=\linewidth,angle=0] {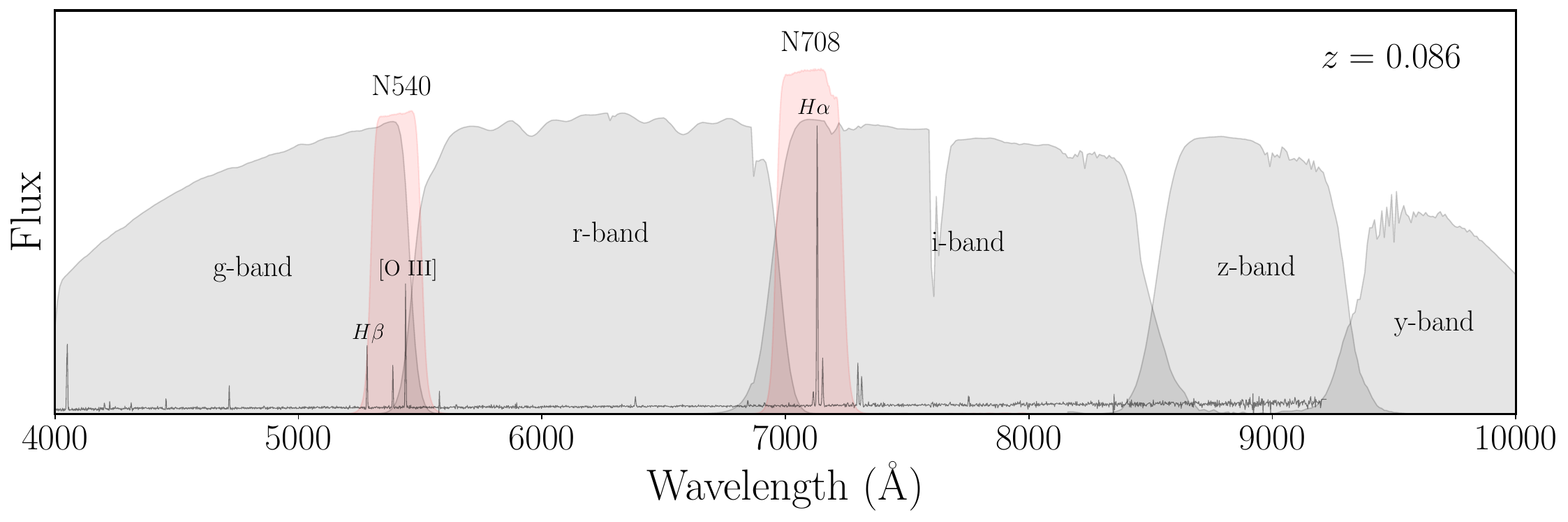}
\caption{Broadband HSC \textit{grizy} filters and medium-band Merian N540 and N708 filters shown over a star-forming dwarf galaxy spectrum from SDSS with $z =0.086$. The Merian N540 and N708 filters are designed to cover [\ion{O}{3}] and H$\alpha$ emission in the redshift range $0.06<z<0.1$. The N708 filter overlaps completely with the HSC $i$-band filter and partially with the HSC $r$-band filter. \label{fig:filters_and_spec}}
\end{center}
\end{figure*}

By measuring and correcting for the stellar continuum in the galaxies, we can use the medium-band Merian filters to map and measure H$\alpha$ emission and compare its distribution to that of the stellar light. With our large sample, we can conduct a statistical analysis of the map morphologies, using nonparametric morphological parameters – simple geometric measures of a galaxy's light distribution that do not rely on assumptions of underlying models. Previous studies have applied this analysis to broadband imaging to identify mergers \citep{Lotz2006}, to compare simulations to observations \citep{Rodriguez-Gomez2019}, and to investigate the physical processes driving star formation in galaxies \citep{Yesuf2021, Bottrell2023}. So far, no studies have extended this analysis to H$\alpha$ maps of nonlocal galaxies, and particularly not to a large, consistently processed sample of such galaxies probing the low-mass regime. 

In this paper, we aim to do exactly that, using the unique Merian data products to create and analyze H$\alpha$ maps of a large sample of low-mass galaxies at $0.064<z<0.1$. In Section~\ref{sec:data} we describe the Merian subsample we use in this work, namely the set of sources from the first Merian data release with available spectroscopy and measured spectroscopic redshifts. In Section~\ref{sec:continuum}, we present a novel method for estimating the contribution of the stellar continuum through the Merian N708 medium band and use this estimated value to construct maps of H$\alpha$ emission. In Section~\ref{sec:morphology}, we detail and calculate the nonparametric morphological statistics used in our analysis. In Section~\ref{sec:results}, we present our findings – the main trends in continuum and H$\alpha$ morphology with physical parameters. Finally, in Section~\ref{sec:discussion}, we consider how these results inform our understanding of galaxy-scale star formation, particularly for low-mass galaxies. 

Throughout this paper, we adopt a flat $\Lambda$CDM model with $H_0=70$ km s$^{-1}$ Mpc$^{-1}$ and $\Omega_m=0.3$.


\section{Data from the Merian Survey} \label{sec:data}

Our primary dataset for this study is the seven-band imaging data from the Merian survey data release 1 (DR1; Danieli et al. 2024, in review). Merian is an optical medium-band imaging survey aiming to cover $\sim800$ $\deg^2$ of the HSC-SSP wide layer with two filters: N540 ($\lambda_c = 5400$\,\AA, $\Delta \lambda = 210$\,\AA) and N708 \citep[$\lambda_c = 7080$\,\AA, $\Delta \lambda = 275$\,\AA,][]{Luo2023}. The two filters, which are installed on the 4\,m Victor M. Blanco telescope at the Cerro Tololo Inter-American Observatory, are designed to detect [\ion{O}{3}] and H$\alpha$ emission for galaxies at $0.06\lesssim z\lesssim0.1$. Joined by the HSC-SSP \textit{grizy} broadband imaging data \citep{Aihara2018, Aihara2019, Aihara2022}, the Merian catalog provides aperture-matched photometry in the seven bands ($grizy$, N708, and N540) processed jointly using the Rubin Observatory LSST Science Pipelines\footnote{\url{https://pipelines.lsst.io/}}\citep{Bosch2018, Bosch2019} and photometric redshifts to all sources (Danieli et al. 2024, in review; Luo et al. 2024, in preperation). The HSC-SSP and Merian filters are shown along with an example of a dwarf galaxy spectrum in \autoref{fig:filters_and_spec}. 

The first Merian data release (Danieli et al. 2024, in review) covers an area of 234 $\deg^2$ and consists of over 93 million sources, each of which has at least a 5$\sigma$ detection in at least one of the two medium bands. Observations for the Merian survey are ongoing, and DR1 covers only a subset of the final Merian footprint.

In addition to the imaging data, Merian also obtained \merianspecsample\ spectra of dwarf galaxies in the COSMOS field using the Magellan/IMACS (PIs: Ting and Danieli) and Keck/DEIMOS (PIs: Leauthaud and Kado-Fong) spectrographs (Luo et al. 2024, in preparation). The targets for this spectroscopic sample were selected using COSMOS photometric redshifts \citep{Laigle2016} to identify galaxies with $8<\log (M_\star/M_\odot)<9$ at $0<z<0.15$ with typical $r$-band magnitudes of $\sim$22~mag. While Merian provides photometric redshifts to all the sources in the survey footprint, here we exclusively employ sources with preexisting spectroscopy and determined spectroscopic redshifts to eliminate further uncertainty from photometric redshifts.

To select sources with known spectroscopic redshifts, we cross-match the Merian DR1 catalog with the publicly available spectroscopic data from the Galaxy and Mass Assembly survey \citep[GAMA;][]{Baldry2018} and the Sloan Digital Sky Survey \citep[SDSS;][]{Almeida2023} and with the proprietary Merian spectroscopic sample described above (Luo et al. 2024, in preparation). The Merian footprint overlaps with GAMA fields G02, G09, G12, and G15. Using a cross-matching distance of 0.3\arcsec, we find \gamamerianmatches\ sources with a matching GAMA spectrum and \sdssmerianmatches\ with a matching SDSS spectrum. Of these matches, \sdssandgamamerianmatches\ have a matching spectrum in both GAMA and SDSS, but we note that some of the GAMA spectra are drawn from the SDSS sample. We also find \merianmerianmatches\ sources with spectra from the Merian spectroscopic sample. In total, we find \specsamplesize\ Merian DR1 sources with spectroscopic matches and reliable redshift estimates – about 0.2\% of the total number of sources in DR1. For this study, we choose a redshift range of $0.064 < z < 0.1$ – slightly different from that described in \citet{Luo2023} – as various parts of the focal plane do not have full transmission through the N708 filter beyond this redshift window. We find \specsamplesizeinband\ sources with spectroscopic redshifts within this redshift range.

\begin{figure*}[t]
\begin{center}
\includegraphics[width=\linewidth,angle=0] {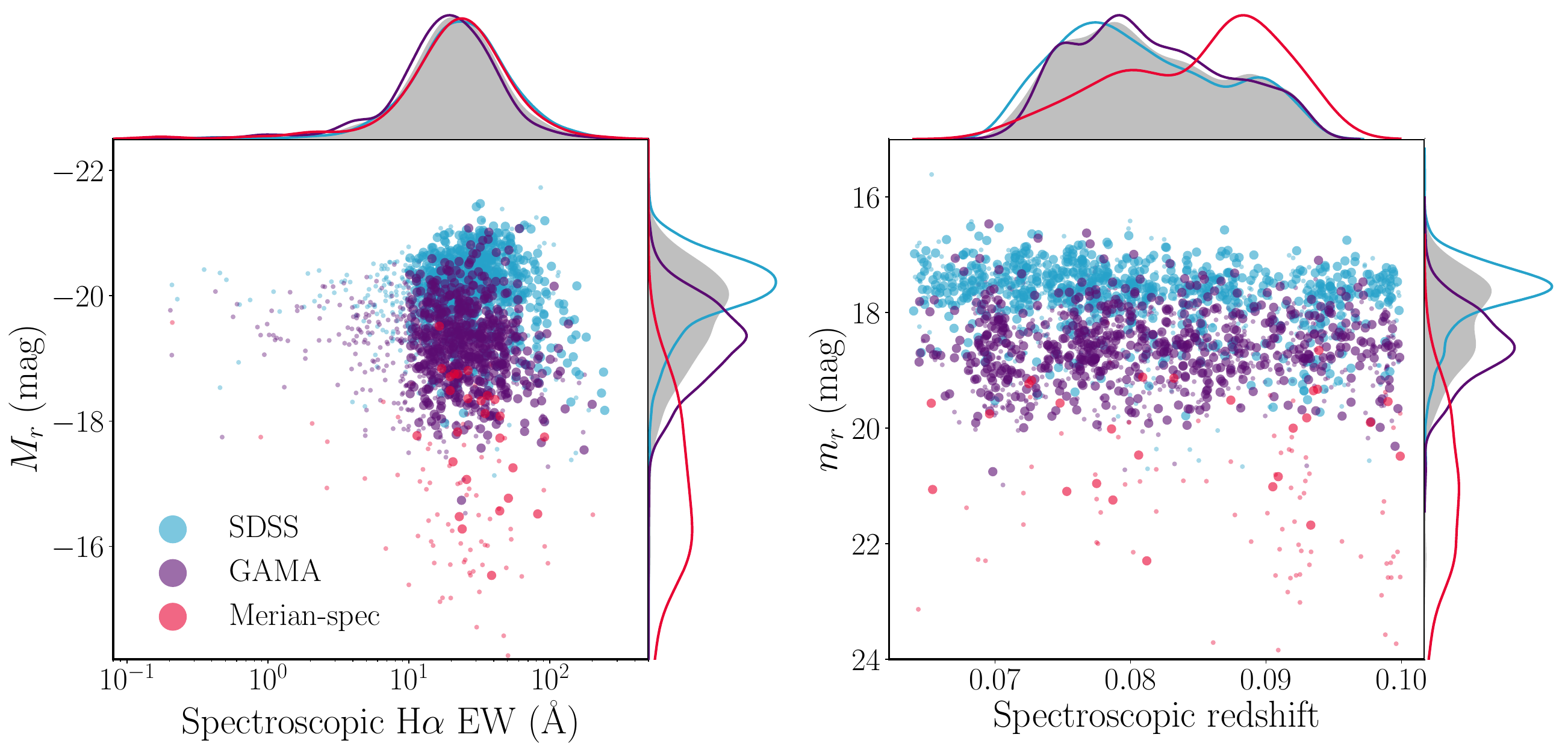}
\caption{Left: distribution of HSC $r$-band absolute magnitude and spectroscopic H$\alpha$ EW for our sample of galaxies in Merian DR1 with preexisting spectroscopy described in Section~\ref{sec:data}, with points colored by the source of the spectra. Smaller points represent galaxies that are not included in our final morphology sample as described in Sections~\ref{subsec:validation} and \ref{subsec:quality}. Marginal distributions are shown for the entire spectroscopic sample in gray and for subsets of the sample based on the source of their spectra in colors matching the scatterplots. Right: the same as the left panel but for HSC $r$-band apparent magnitude vs. spectroscopic redshift.\label{fig:HA_vs_Mr}}
\end{center}
\end{figure*}

While the SDSS and GAMA sources have stellar mass estimates from fits to the sources' spectral energy distributions (SEDs), for consistency we calculate the stellar masses of all the sources in our sample using the HSC $g-r$ color to estimate stellar mass-to-light ratios based on \citet{Into2013} and the HSC $r$-band luminosity to estimate stellar mass. Our masses are in good agreement with those reported by SDSS and GAMA. Because the focus of our study is low-mass galaxies, we include only the \specsamplesizeinbandlowmass\ sources with $\log (M_\star/M_\odot)<10.3$ in our sample. This allows us to compare the dwarfs in our sample to more massive galaxies while avoiding bulge-dominated galaxies \citep{Blanton2009}. Finally, we require that all sources in our catalog have photometry in all seven Merian and HSC bands. This leaves us with a final sample size of \specsamplesizeinbandlowmassallphot. The distributions of some sample properties for the full spectroscopic catalog are shown in \autoref{fig:HA_vs_Mr}. The positions, redshifts, stellar masses, $g-r$ colors, and $r$-band luminosities of the sources in the sample are reported in \autoref{tab:specsample}.

None of these samples are mass complete in the dwarf range in the redshift window of our sample \citep[Luo et al. 2024, in preparation]{Strauss2002, Baldry2010}, but for the purposes of our science, we are interested only in star-forming galaxies and so select sources based on their H$\alpha$ equivalent width (EW; see Section~\ref{subsec:ewfit}). We are therefore strongly biased toward star-forming galaxies by construction and do not expect or require a mass-complete sample.

\begin{deluxetable*}{cccccccc}
\tablecaption{Full Spectroscopic Sample\label{tab:specsample}}
\tablehead{
\colhead{Merian ID} &\colhead{\hspace{0.8cm}R.A.}\hspace{0.8cm}	&\colhead{\hspace{0.8cm}Decl.}\hspace{0.8cm}
&\colhead{\hspace{0.5cm}$z_\textrm{spec}$}\hspace{0.5cm} &\colhead{\hspace{0.5cm}$M_\star$}\hspace{0.5cm} &\colhead{\hspace{0.5cm}$g-r$}\hspace{0.5cm} &\colhead{$L_{r}$}& \colhead{Spec Source}\\
\colhead{}&\colhead{(J2000)}	&\colhead{(J2000)}&\colhead{}&\colhead{$\log(M_\odot)$} &\colhead{(mag)}&\colhead{$\log(L_\odot)$}&\colhead{} \\
\colhead{(1)}&\colhead{(2)}&\colhead{(3)}&\colhead{(4)}&\colhead{(5)}&\colhead{(6)}&\colhead{(7)}&\colhead{(8)}
}
\startdata
2950279968392759821 &	 02:13:26.664 &	 -06:16:37.84 &	 0.0866 &	 9.53 &	0.49 &	 9.44 &	 GAMA \\
2950337142997422666 &	 02:10:41.515 &	 -06:05:18.68 &	 0.0870 &	 9.37 &	0.42 &	 9.39 &	 GAMA \\
2950354735183450022 &	 02:14:20.034 &	 -06:02:02.36 &	 0.0991 &	 9.28 &	0.40 &	 9.33 &	 GAMA \\
2950363531276481619 &	 02:12:52.955 &	 -05:57:21.25 &	 0.0904 &	 9.76 &	0.44 &	 9.75 &	 GAMA \\
2950605423834567700 &	 02:17:57.772 &	 -06:26:24.91 &	 0.0926 &	 9.62 &	0.32 &	 9.78 &	 GAMA \\
2950645006253186410 &	 02:18:03.492 &	 -06:24:22.62 &	 0.0921 &	 9.72 &	0.34 &	 9.85 &	 GAMA \\
2950671394532257761 &	 02:20:03.565 &	 -06:09:30.78 &	 0.0834 &	 9.51 &	0.67 &	 9.18 &	 GAMA \\
2950706578904338273 &	 02:14:31.173 &	 -06:08:12.87 &	 0.0930 &	 9.23 &	0.80 &	 8.74 &	 GAMA \\
2950710976950853296 &	 02:20:13.860 &	 -06:00:27.26 &	 0.0926 &	 9.03 &	0.37 &	 9.12 &	 GAMA \\
2950719773043882287 &	 02:19:00.808 &	 -05:57:50.89 &	 0.0856 &	 9.82 &	0.57 &	 9.63 &	 GAMA \\
\enddata
\tablenotetext{}{\textbf{Note.} The full spectroscopic catalog as described in Section~\ref{sec:data} including \specsamplesizeinbandlowmassallphot\ sources from the Merian photometric catalog with available spectroscopic redshifts.  Column (1): unique identification number assigned to every Merian photometric source. Column (2): R.A. Column (3): decl. Column (4): spectroscopic redshift. Column (5): stellar mass as calculated in Section~\ref{sec:data}. Column (6) $g-r$ color. Column (7) $r$-band luminosity. Column (8): source of spectrum.   (This table is available in its entirety in machine-readable form in the online article.)}
\end{deluxetable*}

\section{Continuum Fitting and H$\alpha$ Maps} \label{sec:continuum}
In order to measure H$\alpha$ emission using the seven bands of Merian and HSC photometry, we must first estimate the contribution from the continuum emission through the N708 filter. In narrowband surveys, this is typically done by assuming a flat continuum in the vicinity of the emission lines and scaling an adjacent broadband filter to correct for the continuum \citep[e.g.,][]{GildePaz2003, Gavazzi2012, Boselli2015,Gavazzi2018, HermosaMunoz2022,Sieben2023}. 

For medium-band surveys, the continuum contribution is more significant and the overlap of the medium-band filters with adjacent broad bands makes the continuum subtraction more complicated. In the Merian survey, the N708 medium-band filter overlaps significantly with the HSC $r$ band and nearly entirely with the HSC $i$ band as seen in Figure~\ref{fig:filters_and_spec}. Consequently, in our redshift range of $0.064<z<0.1$, H$\alpha$ emission (and emission from weaker nearby lines, including [\ion{N}{2}] and [\ion{S}{2}]) necessarily falls into one or both of the adjacent broadband filters. Using the average of the fluxes in the adjacent broadband filters to estimate the continuum under H$\alpha$ will be biased toward overestimating the continuum and underestimating the line emission. Previous medium-band surveys have relied on SED fitting \citep{Terao2022, Chen2023} or derived scaling factors to correct for the intrinsic color in the continuum \citep{Griffiths2021}.

To avoid a biased estimate of the continuum under H$\alpha$ and the H$\alpha$ emission itself, we develop a new continuum fitting method that is less computationally expensive than SED fitting but still utilizes empirical knowledge of the shape of dwarf galaxy continua. 

\subsection{Method Description}\label{subsec:method}

    To estimate the continuum emission through N708, we take advantage of the fact that the continuum in this region of the spectrum can be well approximated by a power law. Because this assumption only holds locally around H$\alpha$, we do not include HSC $g$  or $y$ bands in our fitting. The 4000\,\AA\ break falls into the $g$ band within our redshift range, and so including $g$ band photometry would complicate the shape of the SEDs and invalidate our power law assumption. The $y$ band photometry is shallower and often unreliable for our sources, so we would be disadvantaged if we included it in our modeling. 

    In addition to our power law assumption for the continuum, we also assume that the increase in emission through a photometric band – as compared to the emission we would expect from the continuum only – is a function of the H$\alpha$ EW, [\ion{O}{3}] EW, and redshift. In other words, the increase in emission from continuum only to emission with nebular lines (also including lines other than H$\alpha$ and [\ion{O}{3}], such as [\ion{N}{2}] and [\ion{S}{2}]) is largely determined by the strength of H$\alpha$ and [\ion{O}{3}] and their observed wavelength. We estimate the value of this increase in emission as a scale factor $\eta$, which we derive empirically using a library of simulated spectra and synthetic photometry based on the SED-fitted parameters of low-mass GAMA galaxies and created using the Flexible Stellar Population Synthesis (FSPS) code \citep{Conroy2009, Conroy2010, Johnson2021}. We describe the model in Section~\ref{subsec:model} and the calculation of the scale factor $\eta$ in Section~\ref{subsec:training}.

    We fit the observed photometry using the power-law model and scale factor to estimate the continuum power-law index. With the shape of the continuum, we can calculate its contribution through N708, correct for it, and measure H$\alpha$ EW and flux. Details of this method follow below.

\subsubsection{Model}\label{subsec:model}
    We assume that the continuum itself can be reasonably approximated by a power law of index $a$ anchored at the emission-free flux through the HSC $z$ band:
    \begin{equation}
    F_{\nu, \text{cont}}(\lambda, a)  = \left(\frac{\lambda}{\lambda_z}\right)^a  F^{(z)}_{\nu, \text{cont}}        
    \end{equation}\label{eq:1}
    where $\lambda_z$ is the central wavelength of the HSC $z$ band, $F_{\nu, \text{cont}}$ is the spectral flux density contribution from the continuum alone (i.e. with no line emission), and $F^{(z)}_{\nu, \text{cont}}$ is the continuum flux density integrated through the HSC $z$-band filter. We choose to anchor the power law to the $z$-band value because it is the least contaminated by emission lines in our redshift range. The only emission lines included in FSPS that fall in the $z$ band in our redshift range are [Cl~II] $\lambda$8579\,\AA\ and [C~I] $\lambda$8729\,\AA, both very weak.
    
    We test the power-law assumption using our set of synthetic spectra and photometry described below in Section~\ref{subsec:training} (which spans the parameter space of our data) and find that the continuum fluxes deviate from a power law at the 1\% level for wavelengths between the Merian N540 and HSC $z$ bands.
    
    The continuum flux through a given band, $F^{(x)}_{\nu, \text{cont}}$, can be found by integrating the expression above through the appropriate filter:
    
    \begin{equation}F^{(x)}_{\nu, \text{cont}}(a) = \int_{-\infty}^{\infty} F_{\nu, \text{cont}}(\lambda, a)R_x(\lambda)\ \text{d}\ln\lambda
    \end{equation}\label{eq:2}
    where $R_x(\lambda)$ is the normalized filter response as a function of wavelength and $x$ represents the given filter.

    We can write the flux density with line emission through a photometric band as the flux through that same band without line emission (continuum only) multiplied by a scale factor. We assume that the scale factor can be approximated as a function of H$\alpha$ EW, [\ion{O}{3}] EW, and redshift. For a given filter $x$, we call this scale parameter $\eta^{(x)}({\rm H} \alpha, \textrm{[\ion{O}{3}]}, z)$, and it is defined such that
    \begin{equation}
    F^{(x)}_{\nu} = F^{(x)}_{\nu, \text{cont}}\cdot \eta^{(x)}.
    \end{equation}\label{eq:3}
    We discuss the specifics of how we determine this scale factor in Section~\ref{subsec:training}. We estimate the flux in a given band $(x)$ as a function of H$\alpha$ EW, [\ion{O}{3}] EW, $z$, and $a$ (continuum power-law index):
    \begin{align}
       \hat{F}_{\nu}^{(x)} ({\rm H} \alpha, \textrm{[\ion{O}{3}]}, z, a) = \eta^{(x)} \int\left(\frac{\lambda}{\lambda_z}\right)^a  \frac{F_{\nu}^{(z)}}{\eta^{(z)}}R_x(\lambda)\ \text{d}\ln\lambda.
    \end{align}
    Using this model, we can fix the redshift to the spectroscopic value for the source (which is known by construction for every source in our sample) and fit for the continuum power-law index $a$, H$\alpha$ EW, and [\ion{O}{3}] EW to minimize the log-likelihood of our observed data conditioned on the parameters:
    \begin{align}
    \ln\mathcal{L}({\rm H} \alpha, \textrm{[\ion{O}{3}]}, a)=& -\frac{1}{2}\sum_{x} \left(\frac{F_{\nu}^{(x)} - \hat{F}_{\nu}^{(x)}}{\sigma^{(x)}}\right)^2 \nonumber\\ &+ \ln(2\pi(\sigma^{(x)})^2).
    \end{align}
    
    We do not use the fitted values for H$\alpha$ EW and [\ion{O}{3}] EW and instead use the fitted continuum shape to calculate the line strengths directly from the photometry. The fitted values, derived from FSPS, are highly dependent on the choice of models for stellar evolution, which produce different amounts of ionizing flux and are not well settled \citep{Byler2017}. With the fitted value of $a$, we can estimate the continuum through N708 and measure the H$\alpha$ EW as
    \begin{equation}
    {\rm EW}_{{\rm H} \alpha, 0} = \frac{\int R_{\text{N708}}(\lambda) \ \text{d} \lambda }{R_{\text{N708}}({\rm H} \alpha)} \ \frac{F_{\nu}^{(\text{N708})} - F_{\nu, \text{cont}}^{(\text{N708})}}{F_{\nu, \text{cont}}^{(\text{N708})}}
    \end{equation}
    with $F_{\nu, \text{cont}}^{(\text{N708})}$ calculated following Equation 2.  
    
    Lastly, to isolate the H$\alpha$ emission, we need to correct for weaker emission lines that fall within the N708 filter. In particular, this includes [\ion{N}{2}] ($\lambda\lambda$ 6548, 6584) and [\ion{S}{2}] ($\lambda\lambda$ 6717, 6732). We estimate this correction using an empirically derived redshift-dependent factor $c(z)$, described in \autoref{app:emission}. Hence, the corrected EW$_{{\rm H} \alpha}$ is
    \begin{equation}
    {\rm EW}_{{\rm H} \alpha, {\rm obs}} = \frac{1}{c(z)}{\rm EW}_{{\rm H} \alpha, 0}.
    \end{equation}
    Lastly, we convert our EW values to the rest frame:
    \begin{equation}
    {\rm EW}_{{\rm H} \alpha, {\rm rest}} = \frac{{\rm EW}_{{\rm H} \alpha, {\rm obs}}}{1+z} .   
    \end{equation}
    Going forward, we refer to these corrected, rest-frame values as the H$\alpha$ EW.

    \subsubsection{Estimating the Scale Factor}\label{subsec:training}
        
    The remaining question is how to approximate the scale factor $\eta({\rm H} \alpha, \textrm{[\ion{O}{3}]}, z)$ -- i.e. how much to increase the integrated continuum flux to approximate the integrated flux with line emission through each of our Merian and HSC filters. This is a nontrivial problem that is complicated by the variety of continuum shapes and metallicities of dwarf galaxies. 
    
    To incorporate our empirical knowledge of this parameter space for dwarf galaxies, we build a lookup table of simulated spectra and photometry with FSPS based on the SED fits to low-mass GAMA galaxies. We use the Padova stellar evolution isochrones \citep{Girardi2000, Marigo2007, Marigo2008}, which do not include the effects of stellar rotation but have been shown to produce fewer ionizing photons than the MIST isochrones \citep{Choi2016}, which can overestimate line emission as compared to observations \citep{Byler2017}. We also use the MILES stellar spectra library \citep{SanchezBlazquez2006}, a parametric $\tau$-model for the star formation history, a Kroupa initial mass function \citep{Kroupa2001}, and a \citet{Calzetti2000} attenuation law. 

    We query all galaxies from GAMA DR3 with $8< \log (M_\star/M_\odot)<10$ and high-quality SED fits (for details on the SED fitting see \citet{Taylor2011}). For each galaxy, we use the derived quantities of metallicity, $\tau$ (SPS $e$-folding time for an exponentially declining star formation history), galaxy age, and dust obscuration to create two synthetic spectra using FSPS: one with and one without line emission. To span a wide range of H$\alpha$ EWs, we generate additional spectra by artificially perturbing the age and star formation history of the GAMA galaxies. We randomly choose a galaxy age between 4 and 13.7 Gyr and $\tau$ between 1 and 7 Gyr, and we randomly draw 20 redshifts from our range $0.064<z<0.1$. The synthetic galaxies cover a broad range of H$\alpha$ EW, [\ion{O}{3}] EW, and power-law index:  $0{\rm\,\AA} < {\rm EW}_{{\rm H} \alpha} < 450{\rm\,\AA}$, $0{\rm\,\AA} < {\rm EW}_{[{\rm O~III}]} < 350{\rm\,\AA}$, and $-0.8 < a < 3.3$.
    
    We fit a power law to the synthetic emission-free spectrum and integrate this fitted continuum through the Merian and HSC filters. We also integrate the synthetic spectra with emission to obtain emission-free and emission-full photometry ($F^{(x)}_{\nu, \text{cont}}$, $F^{(x)}_{\nu}$) for each of our seven bands.  Lastly, we calculate the EW of H$\alpha$ and the combined EW of the [\ion{O}{3}] $\lambda\lambda$4959 and $\lambda\lambda$5007 lines for each simulated spectrum.

        With the full training set in hand, we create an interpolated mapping: 
        \begin{equation}
        \eta^{(x)}({\rm H} \alpha, \textrm{[\ion{O}{3}]}, z) = \frac{F^{(x)}_{\nu}({\rm H} \alpha, \textrm{[\ion{O}{3}]}, z)}{F^{(x)}_{\nu, \text{cont}}({\rm H} \alpha, \textrm{[\ion{O}{3}]}, z)}
        \end{equation}
        Essentially, this linear interpolation smooths over the diversity in relative strengths of emission lines other than H$\alpha$ or [\ion{O}{3}] and describes how these relative strengths impact the measured photometry as a function of redshift.

    \subsubsection{H$\alpha$ Maps}\label{subsec:ha_maps}
    
    Having estimated the continuum shape for a given galaxy, we can use our fitted continuum power-law indices to construct two-dimensional maps of H$\alpha$ emission for the galaxies in our sample. We estimate the two-dimensional continuum distribution by scaling the HSC $z$-band image by the ratio of the catalog-level continuum flux through N708 to the flux through the $z$ band: $F_\text{cont}^{(\text{N708})}/F^{(z)}$. We choose to scale the $z$-band image as our continuum estimate because, out of the reliable HSC broad bands, it has the lowest emission-line contamination in our redshift range. It is therefore a reasonably stable representation of the uncontaminated emission from the stellar continuum. 
    
    After aligning the $z$-band and N708 images and matching the point spread functions (PSFs), we subtract the scaled $z$-band image from the N708 image to obtain a map of the H$\alpha$ emission. 
    
    We generate a segmentation map for the $z$-band image using the Python implementation of SExtractor \citep{Bertin1996, Barbary2016}, with slight modifications to the default values (a slightly higher minimum contrast ratio for deblending). The maps are used both to identify the pixels associated with the galaxy and to identify pixels belonging to nearby or otherwise contaminating objects. We mask all pixels in the H$\alpha$ map associated with contaminants and measure H$\alpha$ emission through apertures of various sizes and of the whole galaxy (i.e. summing over all pixels in the segmentation map identified as belonging to the primary galaxy). The H$\alpha$ maps themselves are quite irregular, so generating a segmentation map directly from these images would not capture the full extent of a galaxy's flux. 
    
    The H$\alpha$ flux through a given aperture can be calculated from the images as
    \begin{equation}
    F_{{\rm H} \alpha} = \frac{F_{\rm in\ aperture}}{R_{\rm N708}({\rm H}\alpha) \cdot c(z)}
    \end{equation}
    where $F_{\rm in\ aperture}$ is the continuum-subtracted N708 flux, $R_{\rm N708}({\rm H}\alpha)$ is the response function of N708 at the wavelength of H$\alpha$ and $c(z)$ is the correction for additional line contamination as described in \autoref{app:emission}.

\subsection{Validation}\label{subsec:validation}
 \begin{figure*}[t]
\begin{center}
\includegraphics[width=0.48\linewidth,angle=0]{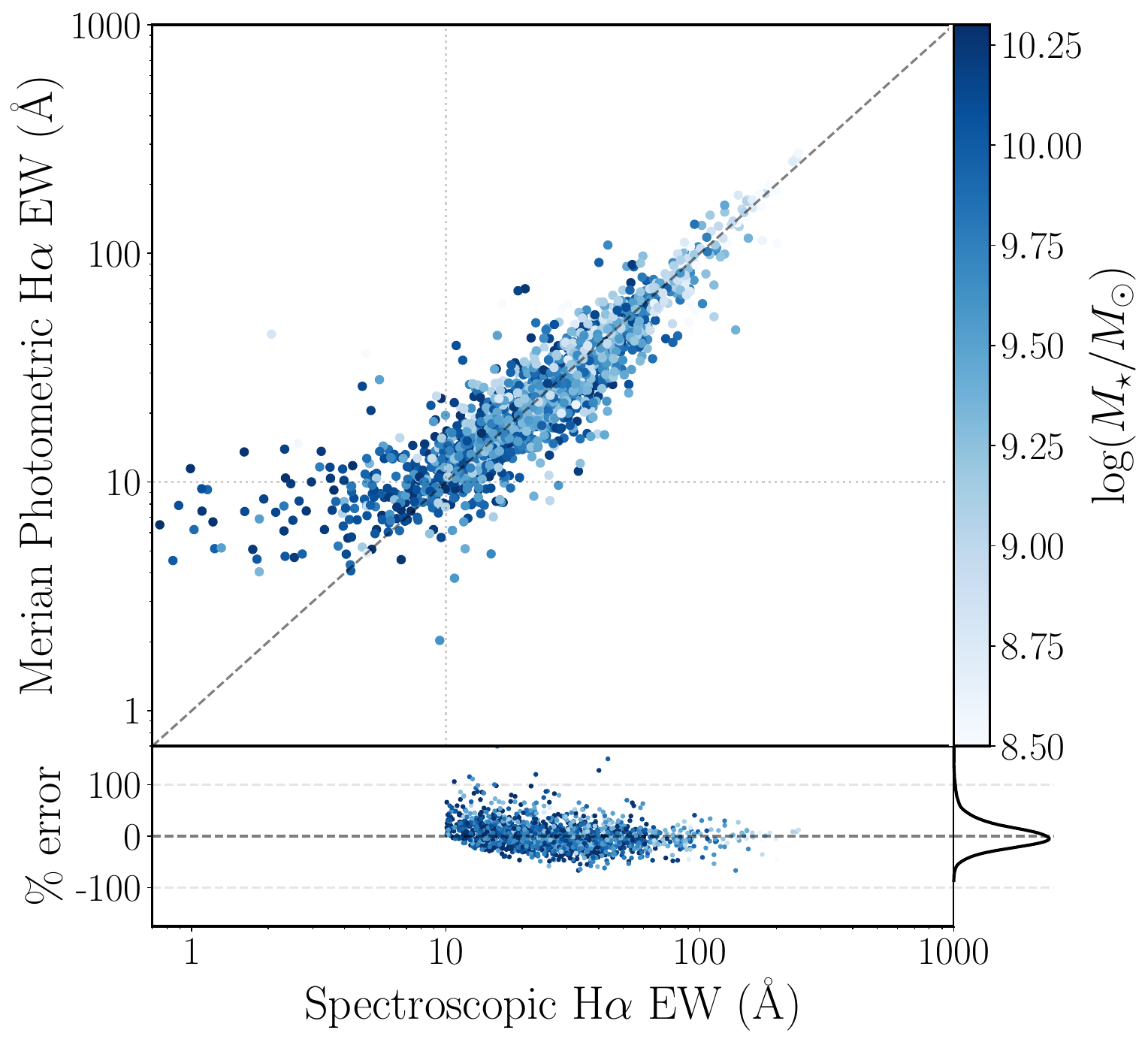}
\includegraphics[width=0.48\linewidth,angle=0]{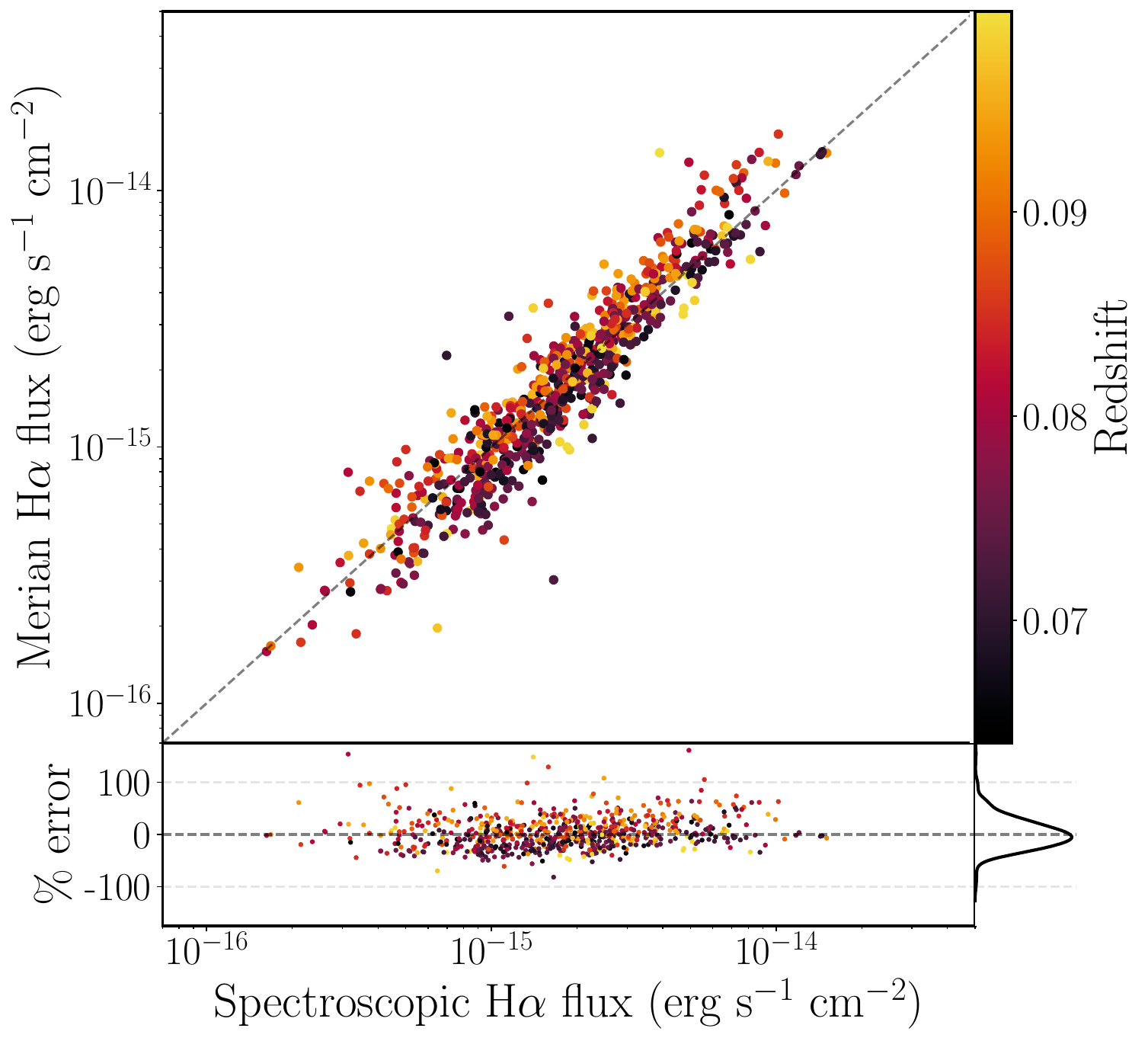}
\caption{Left: values of H$\alpha$ EW estimated from the Merian photometry as described in Section~\ref{subsec:method} compared to spectroscopic values, colored by stellar mass for all sources in the spectroscopic sample with continuum fits. The bottom panel shows the percent error of the fitted values compared to the spectroscopic values as a function of spectroscopic H$\alpha$ EW, excluding sources with H$\alpha$\,EW\,$<10$\,\AA. The bottom panel also shows the marginalized distribution of errors plotted to the far right. We see that the error in the fitted H$\alpha$ EW increases significantly for weaker systems, which is why we focus only on star-forming galaxies and impose a minimum value of 10\,\AA. Right: the same as the left panel, but comparing the H$\alpha$ flux integrated through a 3\arcsec\ aperture from the H$\alpha$ maps described in Section~\ref{subsec:ha_maps} to the H$\alpha$ flux from the SDSS spectra taken through a 3\arcsec\ fiber. This plot includes only sources with spectroscopic matches from SDSS. The points are colored by redshift, with no significant trend observed.   \label{fig:catalogew}}
\end{center}
\end{figure*}
\subsubsection{Fitting to Synthetic Data}
As a first test, we ran the fitting procedure on the synthetic photometry described in Section~\ref{subsec:training}. We only used the synthetic photometry created using the best-fit GAMA star formation histories reported from the SED fitting with a uniform uncertainty of 1\% \citep{Taylor2011}. The synthetic data have $45{\rm\,\AA} < {\rm EW}_{{\rm H} \alpha} < 315{\rm\,\AA}$, $3{\rm\,\AA} < {\rm EW}_{[{\rm O~III}]} < 250{\rm\,\AA}$, and $-0.2 < a < 3.3$. We recovered the power-law index, H$\alpha$ EW, and [\ion{O}{3}] EW to within 5\%.

\subsubsection{Fitting to Merian Photometry}\label{subsec:ewfit}

To further validate our fitting method, we estimated values for the power-law index and calculated the H$\alpha$ EW for each of the galaxies in our spectroscopic sample described in Section~\ref{sec:data} following the method presented above. The measured photometric H$\alpha$ EWs are reported along with the spectroscopic values of H$\alpha$ EW for the subset of the sample described in Section~\ref{subsec:quality} in \autoref{tab:morphsample}.

\begin{figure*}[t]
\begin{center}
\includegraphics[width = 0.9\linewidth]{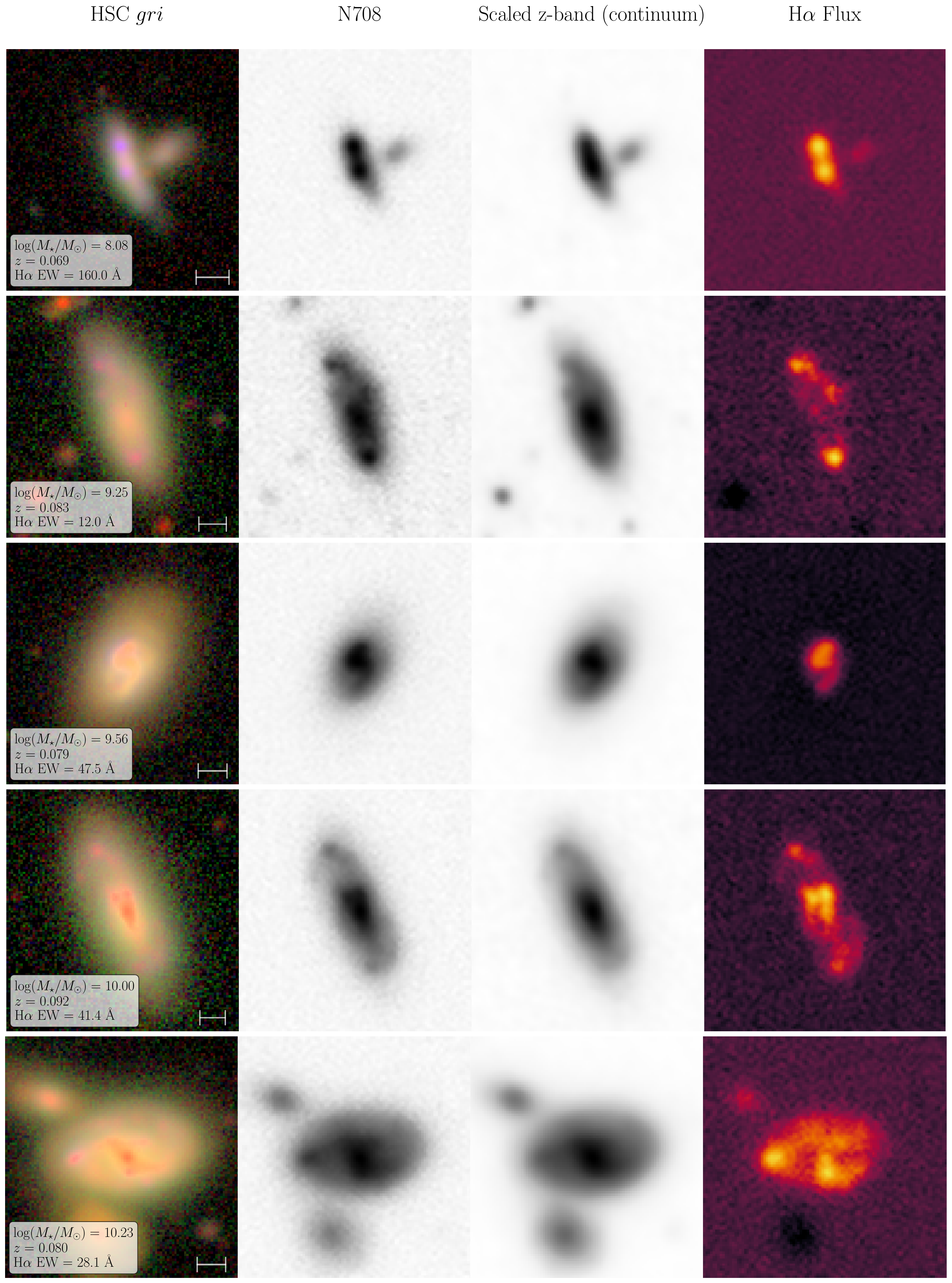}
\caption{Each row shows (from left to right) an HSC-$gri$ color image; the Merian N708 cutout; the scaled $z$-band image, which is our estimate for the two-dimensional continuum distribution; and the H$\alpha$ map. The H$\alpha$ map is the difference between the N708 image and the continuum image shown. The galaxies displayed were randomly drawn from each stellar mass quintile and increase in mass with descending rows. The stellar mass, redshift, and H$\alpha$ EW calculated from the catalog photometry are displayed in the leftmost panel. The bar in the leftmost panel shows the scale of 3 kpc at the redshift of the source. \label{fig:ha_map_grid}}
\end{center}
\end{figure*}

We find that our fitting is generally unreliable (with a median error greater than 80\%) for sources with H$\alpha$\,EW\,$<10$Å where the effects of stellar absorption become significant, so we restrict our comparison to sources with both spectroscopic and fitted photometric EWs above this value. Additionally, \failedfit\ of our sources did not yield reliable continuum fits, either because of poor-quality photometry in one of the five bands used in the continuum fitting or because the H$\alpha$ emission was too low and therefore outside the bounds of our training set (in some cases, there appeared to be evidence of H$\alpha$ absorption). This cut represents a selection based on specific star formation rate (SSFR), with more high-mass galaxies removed from our sample than low-mass galaxies owing to the mass dependence of SSFR. As we are only interested in star-forming galaxies in our analysis, this cut will not impact trends with SSFR that we consider below. 

We compare the fitted H$\alpha$ EWs measured using our procedure to the spectroscopic values from GAMA and SDSS in the left panel of Figure~\ref{fig:catalogew}. We find that the error distribution is fit well by a Gaussian with $\mu = -2\%$ and $\sigma=28\%$. 

We also compare the fitted power-law indices to those we measure from matched SDSS spectra and found that the errors were well fit by a Gaussian with $\mu = -5\%$ and $\sigma=16\%$. Overall, we are able to recover the spectroscopic power-law indices and H$\alpha$ EWs with a high level of confidence. 

We note that, for the purpose of direct comparison, we do not account for stellar H$\alpha$ absorption in our values of H$\alpha$ EW as shown in \autoref{fig:catalogew}. The reported values of H$\alpha$ EW from GAMA do not include such a correction, and values with and without an absorption correction are available for SDSS. For a fair comparison, we use values from all sources without a correction for stellar absorption. Without a full fit of the stellar continuum model, it is not possible to correct properly for the underlying absorption on an object-by-object basis, but from our set of synthetic spectra we find that the EW of the H$\alpha$ absorption is $\sim2.5$\,\AA\ on average, consistent with typical values reported in the literature \citep{Meurer2006}. While some studies choose to implement a standard correction of $\sim1-2$\,\AA\ \citep{GildePaz2003, Meurer2006, Gavazzi2012, Hopkins2013}, the impact of this correction is less significant for star-forming galaxies, whose continuum is dominated by young O- and B-type stars. Given that our sample is selected to have H$\alpha$\,EW\,$> 10$\,\AA, we choose not to implement this correction. Such a correction could increase the values of the SFR we calculate below by up to 20\%, but closer to 1-5\% on average. Given that our conclusions are based on trends with SFR and not on the absolute values of the SFR alone, such a correction would not affect our results.

We also note that our continuum fitting method is limited by the parameter space in our lookup table for estimating the scale factor $\eta$ as described in Section~\ref{subsec:training}. Galaxies with much higher mass, much lower SFR, or much lower metallicities than our training set would likely not be well fit by our current lookup table. The parameter space of the lookup table could be extended to a broader range of stellar mass.

\subsubsection{Flux from H$\alpha$ Images}

We create H$\alpha$ images for the galaxies in our sample as described in Section~\ref{subsec:ha_maps}. Examples of H$\alpha$ maps for galaxies of various masses are shown in \autoref{fig:ha_map_grid}. We compare the H$\alpha$ fluxes measured through a fixed circular aperture with a 3\arcsec\ diameter to the reported H$\alpha$ fluxes in the SDSS catalog, which were measured through a 3\arcsec\ fiber. 

We note that after visual inspection \artifacts\ sources out of \samplesizegoodfit\ with good continuum fits and H$\alpha$\,EW\,$> 10$\,\AA\ were found to have artifacts overlapping with the galaxy and were therefore removed from subsequent analysis. Similarly, \staroverlap\ sources were removed owing to the contaminating presence of an overlapping foreground object.

We compare our measured flux values to the SDSS flux values in the right panel of \autoref{fig:catalogew}. We find that our measured fluxes are in strong agreement with the reported spectroscopic values, with an average scatter of $\sim$ 28\%. 
The H$\alpha$ luminosities measured through a 3\arcsec\ aperture and for the galaxy as a whole as described in Section~\ref{subsec:ha_maps} are reported in \autoref{tab:morphsample}.

\begin{figure*}[t]
\begin{center}
\includegraphics[width = 1\linewidth,angle=0]{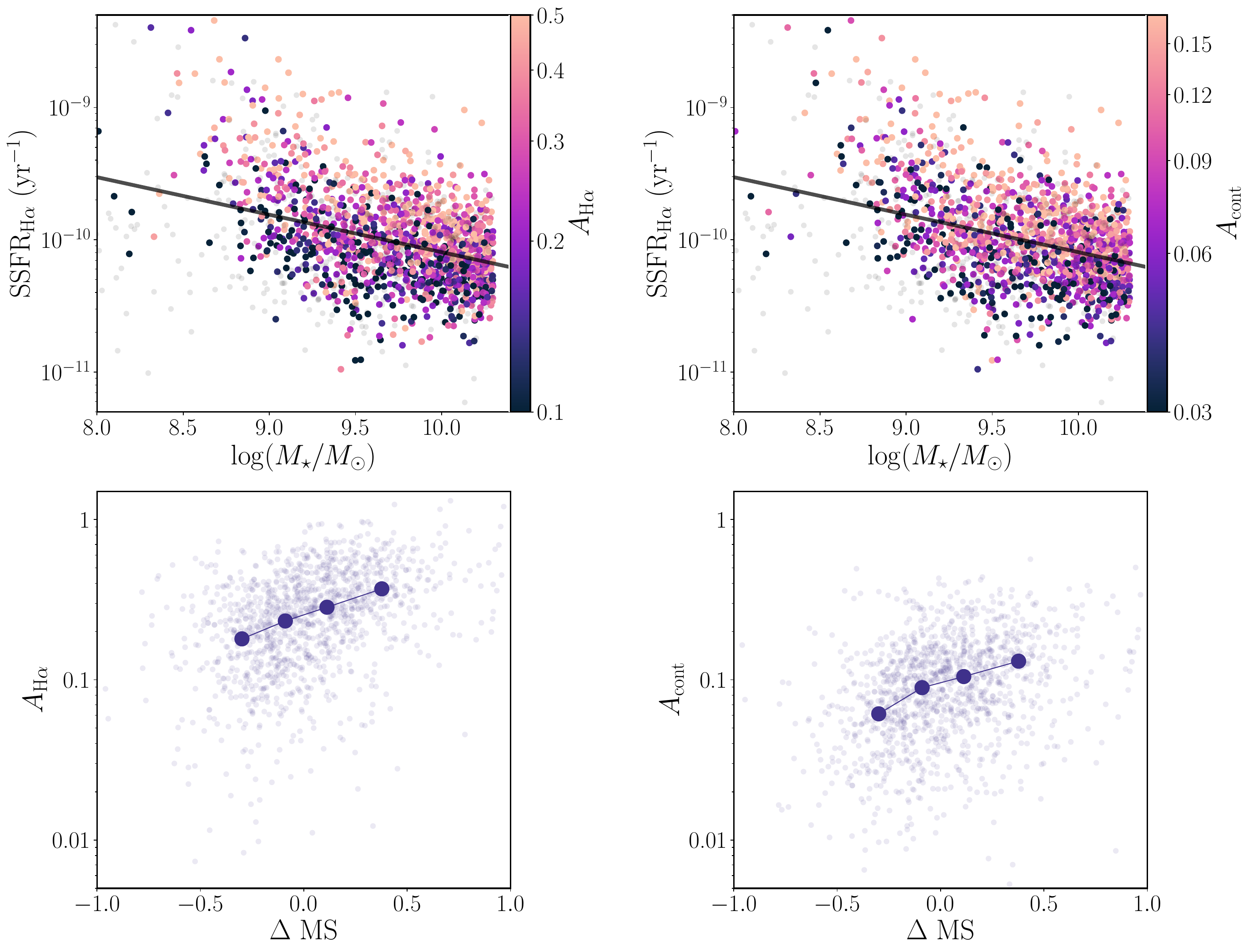}
\caption{Top left: the SSFR derived from the H$\alpha$ flux vs. the stellar mass of our sources as calculated in Section~\ref{sec:data} colored by the asymmetry parameter of the H$\alpha$ maps ($A_{{\rm H}\alpha}$). Light-gray points indicate sources with reliable SSFR measurements that were used to compute the SFMS but did not yield reliable morphological parameters. Sources above the MS (with higher $\Delta$MS) are typically more asymmetric. Top right: the same as the top left panel, but colored instead by the value of the asymmetry parameter measured from the continuum image ($A_{{\rm cont}}$). The same trend is apparent, with $A_{{\rm cont}}$ increasing with distance above the MS. (Bottom) The bottom two panels show this trend explicitly, plotting distance from the MS ($\Delta$MS) vs. $A_{{\rm H}\alpha}$ and $A_{{\rm cont}}$ respectively, with median values shown for a number of $\Delta$MS bins as opaque circles. Both $A_{{\rm H}\alpha}$ and $A_{{\rm cont}}$ are correlated with distance from the MS. \label{fig:sfms_asym}}
\end{center}
\end{figure*}

\subsection{Measuring SFR}\label{subsec:sfr}
While we did not apply a correction for dust extinction when comparing to similarly uncorrected values in Section~\ref{subsec:validation}, it is necessary to correct for extincted emission before using $L_{{\rm H} \alpha}$ to compute the SFR. We correct for foreground extinction using the SFD dust maps of \citet{Schlegel1998} and \citet{Schlafly2011}. For maximum consistency and to avoid introducing additional significant scatter, we derive a mass-dependent average correction for internal attenuation based on the SED fits to the GAMA sources \citep{Taylor2011}. We found corrections based on the Balmer decrement from the spectra to be unreliable for the GAMA sources. Details of this approach are provided in \autoref{app:dust}.

Using the extinction-corrected total H$\alpha$ flux integrated from the H$\alpha$ maps, we calculate the galaxies' SFR according to \citet{Hao2011, Murphy2011}, and \citet{Kennicutt2012}:
\begin{align}
    \log\text{SFR}_{{\rm H} \alpha} (M_\odot \text{ yr}^{-1}) = \log L_{{\rm H} \alpha} (\text{erg s}^{-1}) - 41.27.
\end{align}
We also calculate the specific star formation rate of the galaxies and show the distribution with $\log (M_\star/M_\odot)$ in the top panels of Figure~\ref{fig:sfms_asym}. We fit a line to our star-forming main sequence (SFMS), which we note is not based on a complete sample and does not represent a thorough computation of the SFMS for the Merian sample. We defer this analysis to future work and use the SFMS only in a relative sense to identify which galaxies are particularly active in star formation relative to our sample.

\section{Morphology}\label{sec:morphology}
    The H$\alpha$ maps of the galaxies in our sample reveal a striking variety in the distribution of H$\alpha$ emission, especially as compared to the stellar continua as seen, for example, in \autoref{fig:ha_map_grid}. This highlights one of the unique advantages of our sample – the ability to isolate the two-dimensional H$\alpha$ emission from the stellar continuum. This allows us to analyze its distribution explicitly and to contrast this distribution with that of the stellar emission in order to investigate how sites of star formation are spatially related to the galaxy's potential at large. Our H$\alpha$ map sample is also uniquely large and so requires a systematic approach for morphological analysis. 
    
    To investigate these properties, we measure nonparametric morphological parameters from the images. Traditional parametric approaches for quantifying or classifying galaxy morphology (e.g., fitting two-dimensional S\'ersic models; \citet{Sersic1968}) do not apply well to the H$\alpha$ maps. Unlike stellar emission, the H$\alpha$ emission is frequently clumpy (even at our limited resolution of $\sim$\,1.2\,kpc\,arcsec$^{-1}$) and does not always contain a smooth component – at least not one that is detected by our imaging. 
    Nonparametric statistics are therefore ideal for these maps, as they can be calculated from any image and can be consistently compared across bands. 

    \subsection{Calculating Morphological Parameters}\label{subsec:morph_params}
    We use the publicly available \texttt{Python} package \texttt{statmorph} \citep{Rodriguez-Gomez2019} to calculate the nonparametric morphology statistics for the H$\alpha$ maps and for the $z$-band images, which we take to be representative of the stellar continuum. However, due to the irregularity of the H$\alpha$ maps, we manually fix the centers and segmentation maps (determined as described in Section~\ref{subsec:ha_maps}) for the H$\alpha$ map calculations to those obtained from the $z$-band images. This enables us to conduct a better comparison of the morphological statistics of the two image types. 

    There are several groups of widely used nonparametric morphological parameters calculated by \texttt{statmorph}. We focused our analysis on the concentration-asymmetry-smoothness \citep[CAS;][]{Conselice2003} and Gini-$M_{20}$ \citep{Abraham2003, Lotz2004} statistics, all of which are described in greater detail in \citet{Rodriguez-Gomez2019}.

    While the CAS parameters are among the most commonly used, they are sensitive to resolution and noise effects. Others have shown that the smoothness statistic especially is unreliable at resolutions below 1000\,pc \citep{Lotz2004, Nersesian2023} and \citet{Povic2015} argued that it is not reliable at all for ground-based data. Our resolution ($\gtrsim$ 1~kpc) is too low to calculate meaningful values, and accordingly we have not included smoothness in our results. The concentration parameter is generally more stable, but we similarly do not include it in our analysis, due to the required assumption of circular symmetry and the different surface brightness limits of the medium-band and broadband images. 
    
    While the asymmetry index ($A$) is also affected by resolution and noise, it is not as sensitive as the smoothness parameter \citep{Lotz2004}. Calculations of the smoothness parameter yielded values that were clearly nonphysical, but we found the asymmetry values to be more reasonable – at least for relative comparison within our sample. We therefore calculate and report the value of the asymmetry index. It quantifies the rotational symmetry of the galaxy and is obtained by rotating the image 180$^\circ$ and subtracting it from the original unrotated image. In order to obtain a reliable value, the asymmetry of the background (calculated from an empty region of the image) is also subtracted off. It is calculated as
    \begin{align}
        A = \frac{\sum_{i,j} | I_{ij} - I_{ij}^{180}|}{\sum_{ij} I_{ij}} - A_\text{bgr}
    \end{align}
    where $I_{ij}$ is the flux value of pixel $i,j$ in the original image, $I_{ij}^{180}$ is the flux value in the rotated image, and $A_\text{bgr}$ is the background asymmetry, calculated from a source-free region of the image. The asymmetry is calculated using all pixels within 1.5 Petrosian radii of the asymmetry center, which is defined as the center that minimizes the asymmetry. Again, because we fix the center for the H$\alpha$ images to that of the $z$ band, we note that the H$\alpha$ asymmetry center is the point that minimizes the $z$-band asymmetry and the H$\alpha$ asymmetry is calculated within 1.5 $z$-band Petrosian radii. Using the true H$\alpha$ asymmetry center results in a lower value of $A$ by definition, but the chosen center is frequently far from the continuum center, due to the irregular H$\alpha$ map morphologies. Aside from this one adjustment, the asymmetry of the stellar continuum and the H$\alpha$ maps are calculated identically following the approach described above.

    The Gini coefficient was originally developed by economists to measure the wealth inequality of a population but has been applied to galaxy morphology in order to quantify the homogeneity of the flux distribution. \citet{Abraham2003} and \citet{Lotz2004} were among the first to adapt the Gini coefficient for astronomical applications, accounting for the unreliability of the statistic at low signal-to-noise ratio and presenting a method for constructing a segmentation map that depends on the Petrosian radius. The Gini coefficient is calculated as
    \begin{align}
        G = \frac{1}{\bar{|X|} n(n-1)} \sum_{i=1}^n (2i-n-1) |X_i|
    \end{align}
    where $X_i$ is the flux of the $i$th pixel sorted in ascending order by brightness and $\bar{|X|}$ is the mean of the absolute values of all fluxes associated with the sources as determined by the $z$-band segmentation map. A Gini coefficient of 1 corresponds to a galaxy with all of its flux in a single pixel, and a Gini coefficient of 0 represents perfectly evenly distributed flux. 

\begin{deluxetable*}{ccccccccccc}
\tablecaption{Final morphology sample\label{tab:morphsample}}
\tablehead{
\colhead{Merian ID}&\colhead{EW$_{{\rm H}\alpha, {\rm phot}}$} &\colhead{EW$_{{\rm H}\alpha,  {\rm spec}}$}	&\colhead{$L_{{\rm H}\alpha,  {\rm tot}}$}&\colhead{$L_{{\rm H}\alpha,  {\rm 3\arcsec}}$} &\colhead{$A_{\rm cont}$}&\colhead{$G_{\rm cont}$}&\colhead{$M_{20,{\rm cont}}$} &\colhead{$A_{{\rm H}\alpha}$}&\colhead{$G_{{\rm H}\alpha}$}&\colhead{$M_{20,{{\rm H}\alpha}}$}\\[0cm]
 \colhead{} &\colhead{(\AA)}&\colhead{(\AA)}&\colhead{$\log(L_\odot)$}&\colhead{$\log(L_\odot)$}&\colhead{} &\colhead{} &\colhead{} &\colhead{} &\colhead{} &\colhead{} \\[-0.1cm]
\colhead{(1)}&\colhead{(2)}&\colhead{(3)}&\colhead{(4)}&\colhead{(5)}&\colhead{(6)}&\colhead{(7)}&\colhead{(8)}&\colhead{(9)}&\colhead{(10)}&\colhead{(11)}
}
\startdata
2950337142997422666 &	 14 &	 15 &	 7.06 &	 6.17 &	0.13 &	 0.45 &	 -1.75 &	 0.11 &	 0.45 &	 -0.78 \\
2950354735183450022 &	 23 &	 30 &	 6.92 &	 6.72 &	0.14 &	 0.50 &	 -1.67 &	 0.10 &	 0.48 &	 -1.27 \\
2950363531276481619 &	 18 &	 22 &	 7.32 &	 6.98 &	0.10 &	 0.50 &	 -1.80 &	 0.28 &	 0.50 &	 -1.36 \\
2950710976950853296 &	 27 &	 33 &	 6.75 &	 6.60 &	0.11 &	 0.46 &	 -1.60 &	 0.20 &	 0.53 &	 -0.87 \\
2950948471462439544 &	 24 &	 25 &	 6.89 &	 6.40 &	0.06 &	 0.48 &	 -1.81 &	 0.00 &	 0.43 &	 -1.29 \\
2950952869508950140 &	 19 &	 14 &	 6.63 &	 6.48 &	0.01 &	 0.45 &	 -1.62 &	 0.16 &	 0.52 &	 -1.39 \\
2951001248020583970 &	 21 &	 28 &	 7.11 &	 6.79 &	0.03 &	 0.46 &	 -1.70 &	 0.12 &	 0.51 &	 -0.91 \\
2951001248020596535 &	 22 &	 27 &	 7.15 &	 6.82 &	0.09 &	 0.51 &	 -1.75 &	 0.08 &	 0.53 &	 -1.57 \\
2951014442160132111 &	 26 &	 34 &	 7.10 &	 6.93 &	0.19 &	 0.48 &	 -1.72 &	 0.15 &	 0.50 &	 -1.21 \\
2951049626532208993 &	 35 &	 35 &	 7.07 &	 6.77 &	0.07 &	 0.49 &	 -1.74 &	 0.10 &	 0.52 &	 -1.23 
\enddata
\tablenotetext{}{\textbf{Note.} The sample of \postmorphsample\ Merian sources with available spectroscopy,  photometric and spectroscopic H$\alpha$ EW greater than 10\,\AA, images free from defects, and well-measured morphological statistics as described in Section~\ref{subsec:quality}. Column (1): unique identification number assigned to each Merian photometric source. Column (2): H$\alpha$ EW measured from Merian photometry as described in Section~\ref{sec:continuum}. Column (3): H$\alpha$ EW reported from spectroscopic measurements. Column (4): H$\alpha$ luminosity measured over the whole galaxy as described in Section~\ref{subsec:ha_maps}. Column (5): H$\alpha$ luminosity measured through a 3\arcsec\ aperture. Column (6): asymmetry of the continuum. Column (7): Gini coefficient of the continuum. Column (8): $M_{20}$ statistic of the continuum. Column (9): asymmetry of the H$\alpha$ emission. Column (10): Gini coefficient of the H$\alpha$ emission. Column (11): $M_{20}$ statistic of the H$\alpha$ emission. (This table is available in its entirety in machine-readable form in the online article.)}
\end{deluxetable*}

    The last morphological parameter we consider is the $M_{20}$ statistic, also presented by \citet{Lotz2004}. $M_{20}$ measures the second moment of the brightest 20\% of the galaxy's flux relative to its total second moment. It is calculated as
    \begin{align}
        M_{20} = \log_{10}\frac{\sum_{i} \mu_{i}}{\mu_{tot}}{\rm ,   while} \sum_{i}I_i <0.2I_{\rm tot}
    \end{align}
    where $I_{\rm tot}$ is the total flux of the pixels associated with the source as determined by the segmentation map and $\mu_{\rm tot}$ is the total second moment:
    \begin{align}
        \mu_{\rm tot} = \sum _{i=1}^n \mu_i = \sum_{i=1}^n I_i \left[ (x_i-x_c)^2 
 + (y_i-y_c)^2\right].
    \end{align}
    The center coordinate ($x_c, y_c$) is found as the point that minimizes $\mu_{\rm tot}$ for the continuum image and, as for the other parameters, is fixed to the $z$-band value for the H$\alpha$ maps. While $M_{20}$ is similar to the concentration parameter, it is more flexible. It does not assume circular symmetry and is more influenced by the brightest regions of the galaxy. It is therefore useful for identifying features like bright nuclei or off-center clusters. 

\begin{figure*}[t]
    \begin{center}
    \includegraphics[width = \linewidth,angle=0]{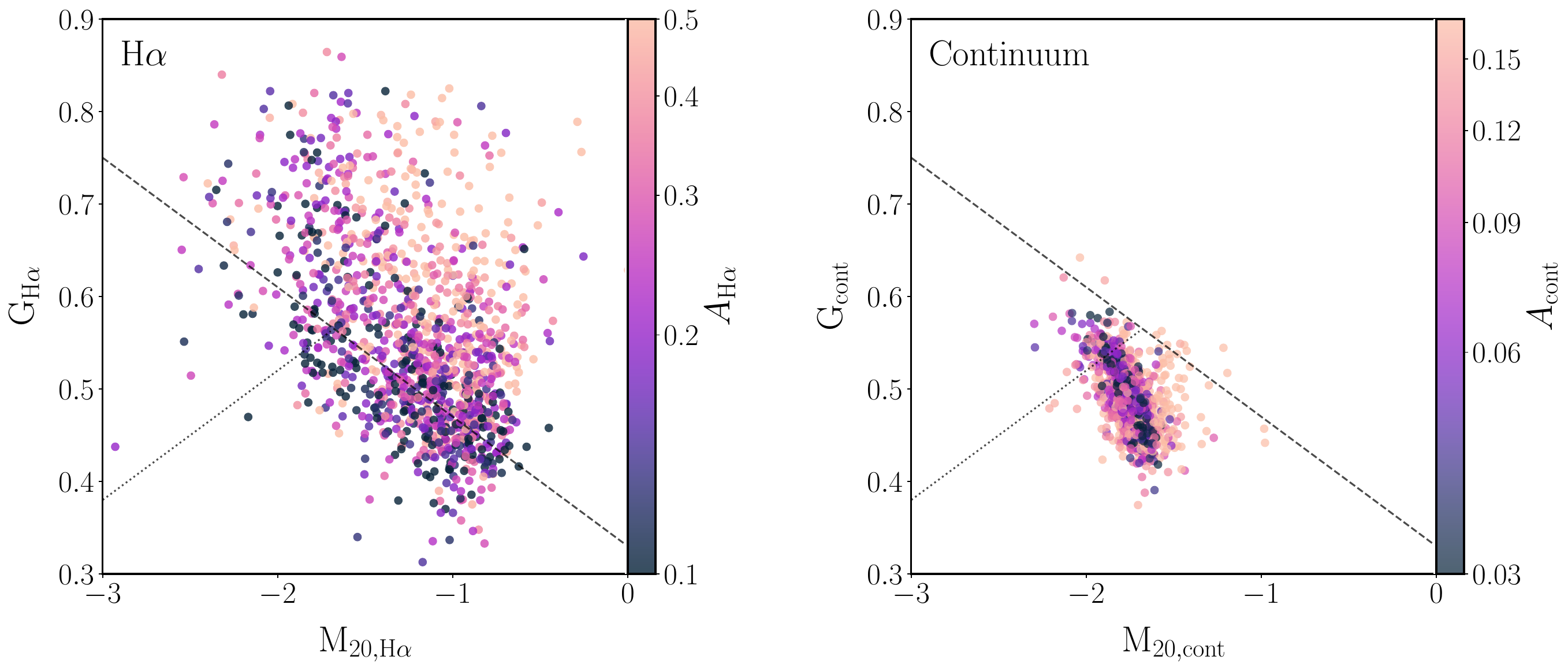}
    \caption{Left: the Gini coefficient vs. the $M_{20}$ statistic for the H$\alpha$ maps colored by H$\alpha$ asymmetry. The dashed and dotted lines are taken from \citet{Lotz2008} and are meant to delineate mergers (above the dashed line) and early- and late-type galaxies (above and below the dotted line, respectively) based on their broadband morphology. Right: the same as the left plot, but with all morphological statistics measured from the continuum images. A small number of galaxies lie above the merger-candidate line; notably, all of these sources have high values of measured asymmetry. The interpretation of the Gini-$M_{20}$ space as presented in \citet{Lotz2008} was derived from broadband morphologies and cannot be directly applied to maps of line emission. While the continuum morphological statistics fall in the expected parameter space, those measured from the H$\alpha$ maps have a much broader distribution; the traditional delineations are not relevant for the H$\alpha$ morphological statistics.   \label{fig:g_v_m20}}
    \end{center}
    \end{figure*}

\begin{figure}[t]
    \begin{center}
    \includegraphics[width = \linewidth,angle=0]{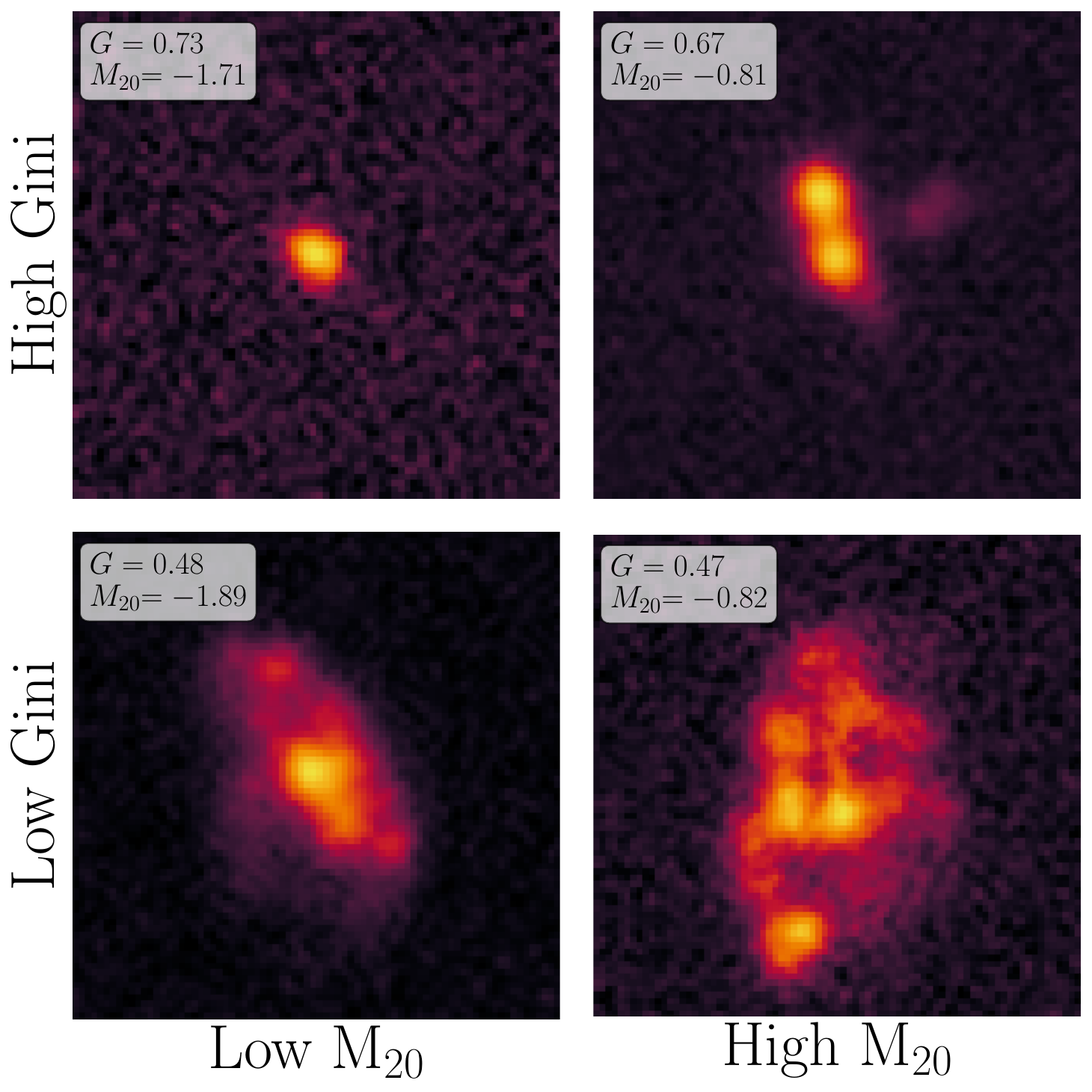}
    \caption{H$\alpha$ maps of four galaxies with extreme values of the Gini coefficient and $M_{20}$ statistic. Maps with high values of $G$ have nonhomogeneous distributions of H$\alpha$ emission, while those with low values of $G$ have their flux more evenly distributed over the galaxy. Maps with low $M_{20}$ have similar second moments for their total distribution and the distribution of their brightest pixels, while those with high $M_{20}$ have second moments that are less similar – i.e., the distribution of their brightest regions does not match the H$\alpha$ distribution at large.   \label{fig:gini_m20_grid}}
    \end{center}
    \end{figure}

    The Gini coefficient and $M_{20}$ statistic are frequently used together to identify mergers or characterize bulge strength \citep{Lotz2004, Lotz2008}. Under this scheme, sources whose broadband morphological statistics fall in the high-Gini, high-$M_{20}$ region of this parameter space are classified as merger candidates. We show the distribution of $G$ versus $M_{20}$ measured from the continuum images in the right panel of \autoref{fig:g_v_m20} along with the merger-separating lines presented in \citet{Lotz2008}. We also show the distribution of the H$\alpha$ parameters in the left panel of this figure, which cover a broader range of the parameter space. We discuss the implications of this difference further in Section~\ref{subsec:comp_ha_cont}. The Gini coefficient and $M_{20}$ statistic are also useful for quantifying the H$\alpha$ homogeneity (or lack thereof) and the degree of central concentration of bright H$\alpha$ clumps. \autoref{fig:gini_m20_grid} shows four galaxies with H$\alpha$ maps with extreme values of $G$ and $M_{20}$ to demonstrate the use of these values in identifying, for example, multiple bright clumps and central nuclei.

    \subsection{Quality Selection}\label{subsec:quality}
    To ensure that we only consider galaxies with reliably calculated morphological parameters, we impose the selection criteria proposed in \citet{Rodriguez-Gomez2019}. We apply the built-in quality selection from \texttt{statmorph}, removing all sources with \texttt{flag > 0} with \texttt{flag} as defined in \citet{Rodriguez-Gomez2019} to indicate a poor-quality measurement. We also remove sources with mean signal-to-noise ratio per pixel below 2.5 and sources with $r_{20}$ (the radius containing 20\% of the galaxy's flux) less than half the FWHM of the PSF. Lastly, we remove any sources that pass all of the above criteria but have unphysical values for asymmetry (i.e. negative). After imposing these selections, we removed \morphremove\ sources, yielding a sample of \postmorphsample\ galaxies with reliable continuum fits, active star formation, good-quality images, and reasonable morphological calculations. We report the values of the asymmetry index, Gini coefficient, and $M_{20}$ statistic measured from the $z$ band and H$\alpha$ images for these sources in \autoref{tab:morphsample}.
    
    The majority of sources removed owing to the morphology quality cut are low-mass (with a median $\log (M_\star/M_\odot)\approx 9.3$), have low surface brightness and low SSFR for their mass, and are physically small. Because they do not have reliable morphological measurements by definition, it is challenging to quantify the bias introduced by this quality cut, but from visual inspection the majority of the sources appear dim and diffuse. To assess the impact of excluding these sources, we analyzed the sample with the ``poor-quality" morphological measurements included and found that it had little effect on our results as described in Section~\ref{sec:results}.
    
    The main impact of this selection is in removing the majority of extremely low-mass sources from our sample – all but one of the 21 sources with $\log (M_\star/M_\odot)< 8$ are removed. As discussed previously, our sample is not complete at the low-mass end, as it is exceedingly difficult to get spectroscopic measurements for these low surface brightness objects. Probing such faint sources is beyond the scope of our data and in fact represents a different physical regime – one characterized by low stellar density and thus less active star formation \citep{Kado-Fong2022}. In the analysis that follows, we restrict our sample to sources with $\log (M_\star/M_\odot) \geq 8$, removing the remaining source below this threshold.

    Additionally, as discussed above, nonparametric morphological parameters can be sensitive to resolution. As our sample spans a range of redshifts, one might be concerned about the impact of variable resolution on subsequent analysis of the galaxies' morphology. Our H$\alpha$ maps have a typical PSF FWHM of $\sim$\,5 pixels, which corresponds to a physical scale of 1.03 kpc at $z=0.064$ and 1.5 kpc at $z=0.1$. Despite the moderate change in resolved physical scale, we find no significant morphological trends with redshift for any of our parameters, providing confidence that any observed trends are representative of intrinsic differences among the galaxies and not simply a result of resolution effects.

    Another challenge is that at lower star formation rates, we may be unable to detect H$\alpha$ emission owing to surface brightness incompleteness, potentially biasing trends between morphology and SFR. To assess the potential significance of this effect, we binned our sources by mass and redshift.  We found that in the high-mass galaxies the lowest SFR surface density pixels are $\sim$1000 times fainter than the highest SFR surface density pixels and we achieve a similar dynamic range in the low-mass galaxies at the same redshift. Hence, we do not believe that the morphological trends in our maps can be attributed to differential incompleteness due to surface brightness.
    

    While we find no mass- or SSFR-dependent impacts of surface brightness limitations, it is likely that some amount of diffuse emission in our galaxies drops below our detection limits. While we have shown that this would have no relative impact on our measurements, it might result in slightly inflated values of the Gini coefficient overall as compared to what would be measured with deeper imaging. 

\subsection{Potential Impact of Dust on H$\alpha$ Morphology}\label{subsec:morph_dust}
Lastly, it is important to consider the potential impact of spatially dependent dust attenuation on our measured morphological parameters for the H$\alpha$ maps. Without high-resolution infrared imaging or integral field spectroscopy, it is not possible to explicitly correct for spatially varying attenuation in our images. Previous work has shown that dust attenuation is typically higher toward the center of galaxies \citep{Boissier2004, Boissier2007, Nelson2016, Kahre2018, Jafariyazani2019}, but has found that both the overall attenuation and the gradient decrease with mass and that there is little attenuation at any radius for the lowest-mass sources ($\log (M_\star/M_\odot)<9.2$). Given that dust attenuation is correlated with metallicity \citep{Boissier2004, Boissier2007} and dwarfs typically do not have significant metallicity gradients \citep{Croxall2009}, it is unlikely that dust would impact the measured morphological parameters of the dwarf galaxies in our sample.

For the higher-mass sources, the gradient in attenuation is potentially more significant, with the H$\alpha$ attenuation reaching up to $\sim$2--3 mag in the central 1 kpc, but declining rapidly to less than 1 mag beyond 1 kpc and becoming negligible beyond 3 kpc \citep{Nelson2016}. We note that this value is an upper limit, as the analysis from \citet{Nelson2016} is based on star-forming galaxies at $z\approx1.4$, which have higher SSFRs (by nearly an order of magnitude) and are dustier than our galaxies at $z<0.1$. 

To approximate the impact of radially dependent attenuation on our measured morphological parameters for the higher-mass galaxies in our sample ($\log(M_\star/M_\odot)>9.8$), we conducted a simplified experiment using the parameterization of H$\alpha$ attenuation from \citet{Nelson2016}, the details of which are presented in \autoref{app:dust_morph}. Ultimately, we find that the asymmetry is not strongly affected by dust but that $G_{{\rm H}\alpha}$ and $M_{20, {\rm H}\alpha}$ could be biased by at most $10-20\%$ for the highest-mass galaxies in our sample. These estimates represent extreme upper limits for the galaxies in our sample, as they are based on galaxies with higher SFRs and our toy model for dust correction is very simplified. These caveats are discussed further in \autoref{app:dust_morph}. A true characterization of the morphological bias would require additional information.

\section{Results}\label{sec:results}

\begin{figure*}[t]
\begin{center}
\includegraphics[width = 0.95\linewidth,angle=0]{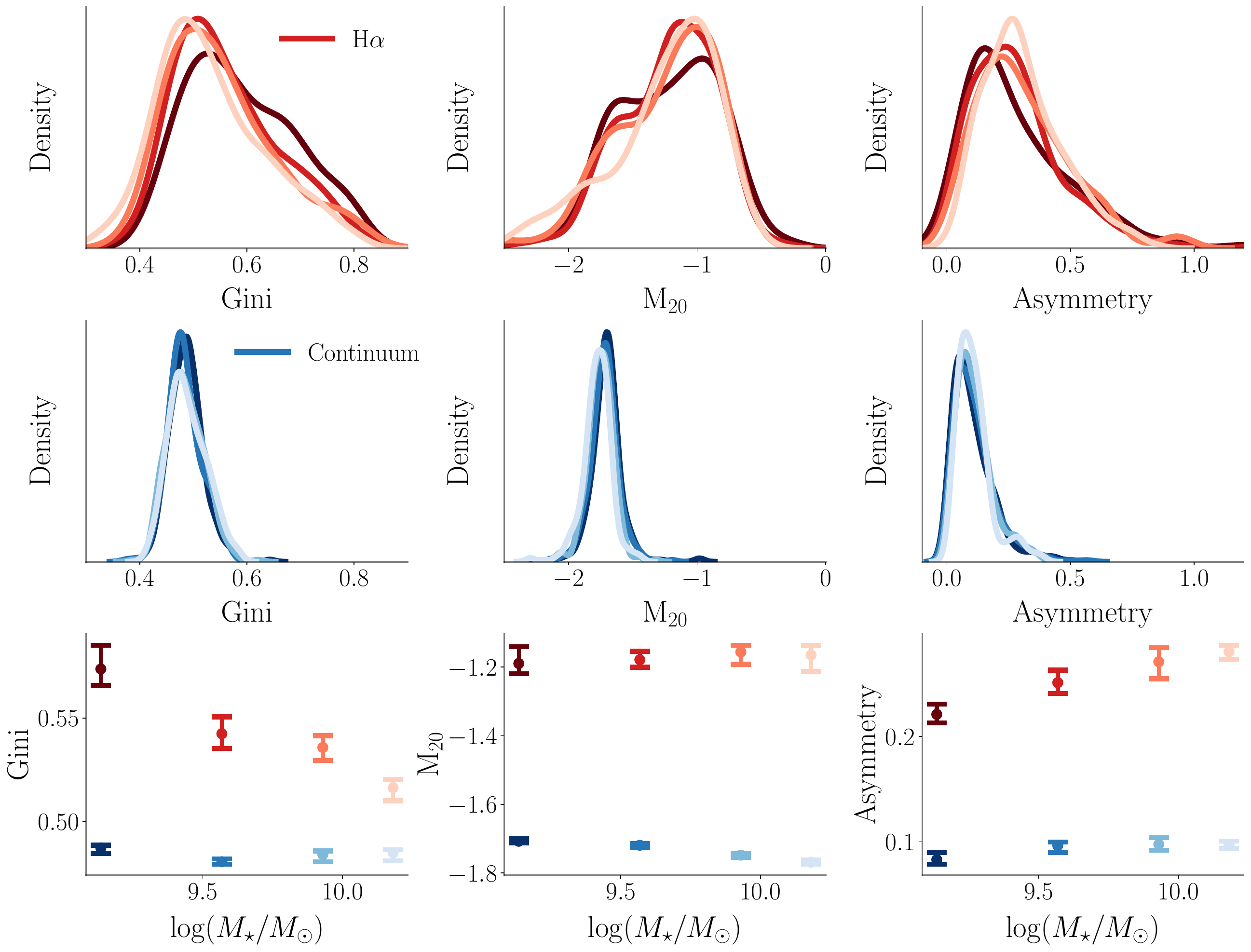}
\caption{Top row: each panel shows the distribution of a given morphological parameter (Gini, $M_{20}$, and asymmetry from left to right) measured from the H$\alpha$ maps and binned by mass. Lighter values of red indicate higher stellar mass. The median mass values in each bin can be read off of the plots in the bottom row. Middle row: the same as the top row, but for the continuum morphological parameters. Lighter-blue values indicate higher stellar mass. The continuum distributions are generally much more tightly distributed than the H$\alpha$ distributions. (Bottom row) The median values of the morphological parameters in each mass bin are plotted as a function of mass. The error bars indicate the 16\% – 84\% confidence intervals on the medians estimated using bootstrap resampling. The red points represent the values for the H$\alpha$ maps, and the blue points show the continuum values. The H$\alpha$ maps are on average less homogeneous, have their brightest regions more extended, and are more asymmetric than the continuum images. The Gini coefficient for the H$\alpha$ maps appears to decrease with mass, with the lowest-mass galaxies having a significantly higher $G_{{\rm H}\alpha}$. There is also a slight trend of decreasing $M_{20, {\rm cont}}$ with mass for the higher mass bins. The lowest-mass galaxies also have more symmetric H${\alpha}$ emission on average. There are no other significant trends with mass observed in the continuum. \label{fig:mass_kdes}}
\end{center}
\end{figure*}

\begin{figure*}[t]
\begin{center}
\includegraphics[width = 0.95\linewidth,angle=0]{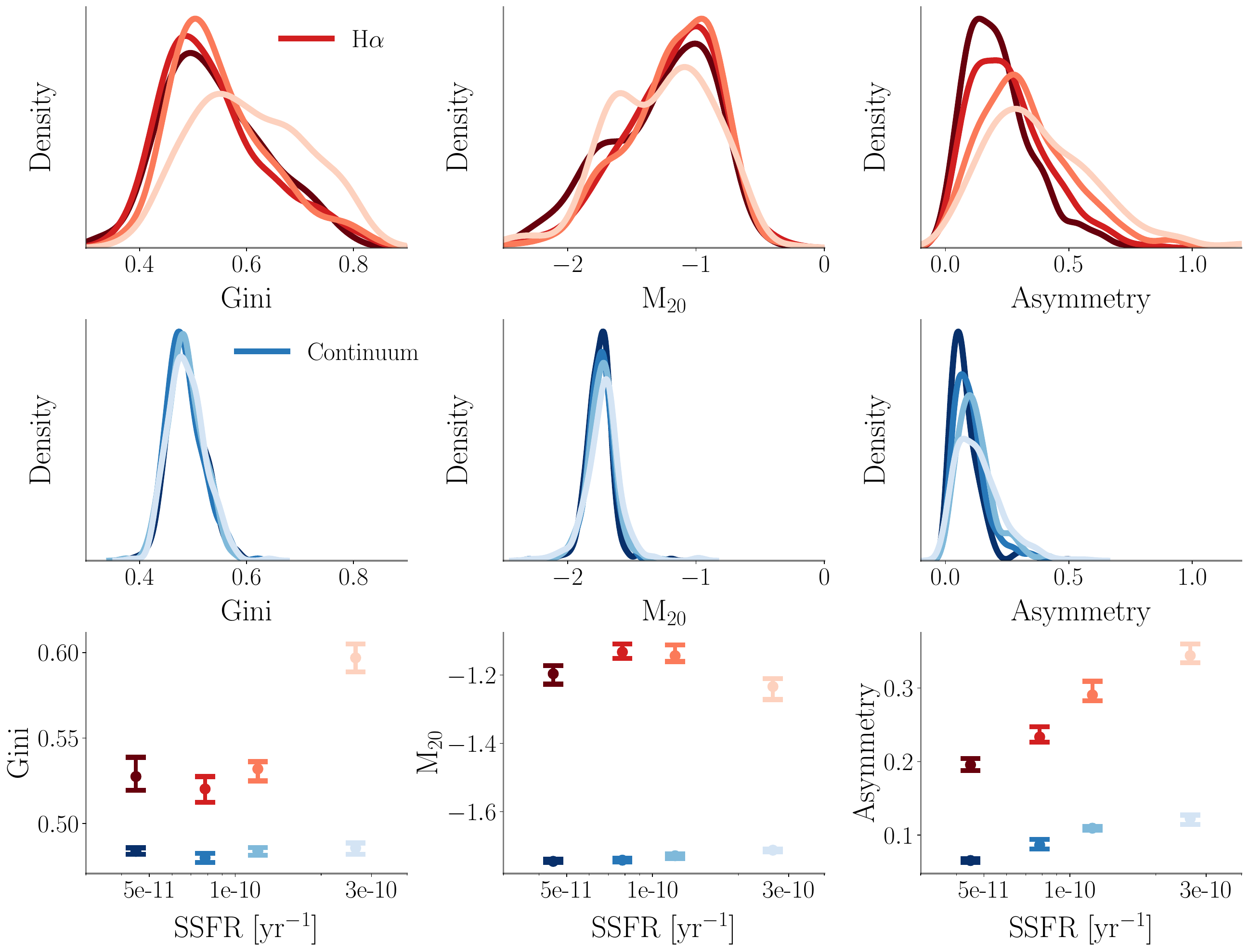}
\caption{Same as \autoref{fig:mass_kdes} but binned by SSFR instead of mass. Asymmetry increases with SSFR, both for the H$\alpha$ maps and for the continua. The highest SSFR bin also has higher values of $G_{{\rm H}\alpha}$ and lower values of $M_{20, {\rm H}\alpha}$ on average, and no such trend is seen in the continua.  \label{fig:ssfr_kdes}}
\end{center}
\end{figure*}

\begin{figure*}[t]
\begin{center}
\includegraphics[width = \linewidth,angle=0]{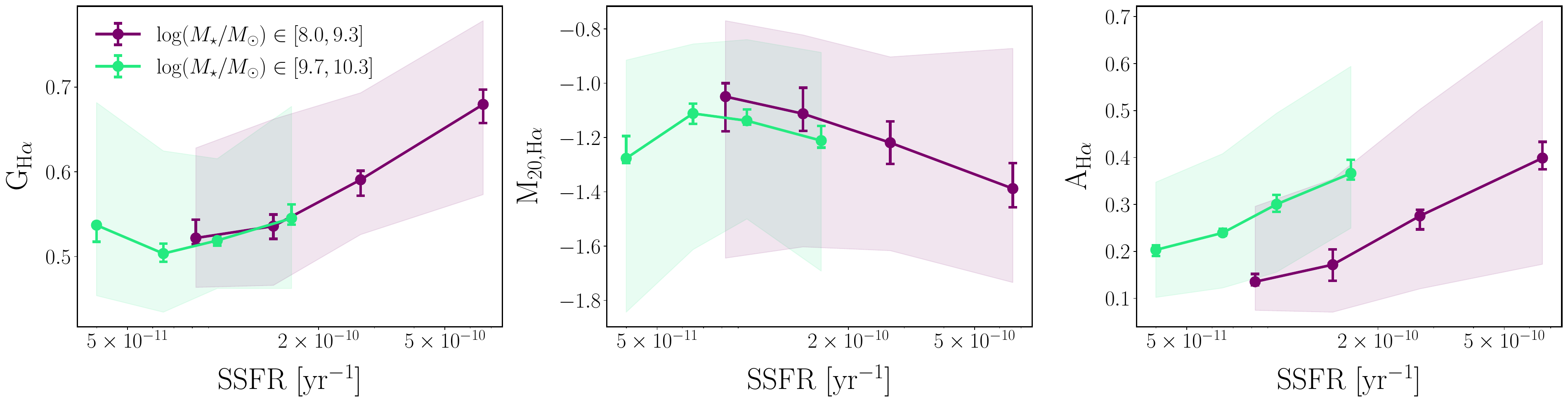}
\caption{Left: the median values of the H$\alpha$ Gini coefficient in bins of SSFR are plotted against SSFR for two different mass bins. The error bars show the 16\%-84\% confidence interval for the median calculated from bootstrap resampling. The shaded region represents the Gini values that contain the central 68\% of points in each bin. We observed no significant trend with SSFR in the Gini coefficient for the higher-mass galaxies, but see an increase with SSFR for the lowest-mass galaxies. Middle: the same as the left panel but for the $M_{20}$ statistic. Again, there is no clear trend for the higher mass bin with SSFR, but the H$\alpha$ maps of the lowest-mass sources tend to have lower $M_{20}$ with higher SSFR. Right: the same as the left panel but for the H$\alpha$ asymmetry index. Both the high- and low-mass bins increase in asymmetry with SSFR, but the low-mass galaxies are less asymmetric overall. \label{fig:gini_asym}}
\end{center}
\end{figure*}

\begin{figure*}[t]
\begin{center}
\includegraphics[width=\linewidth,angle=0]{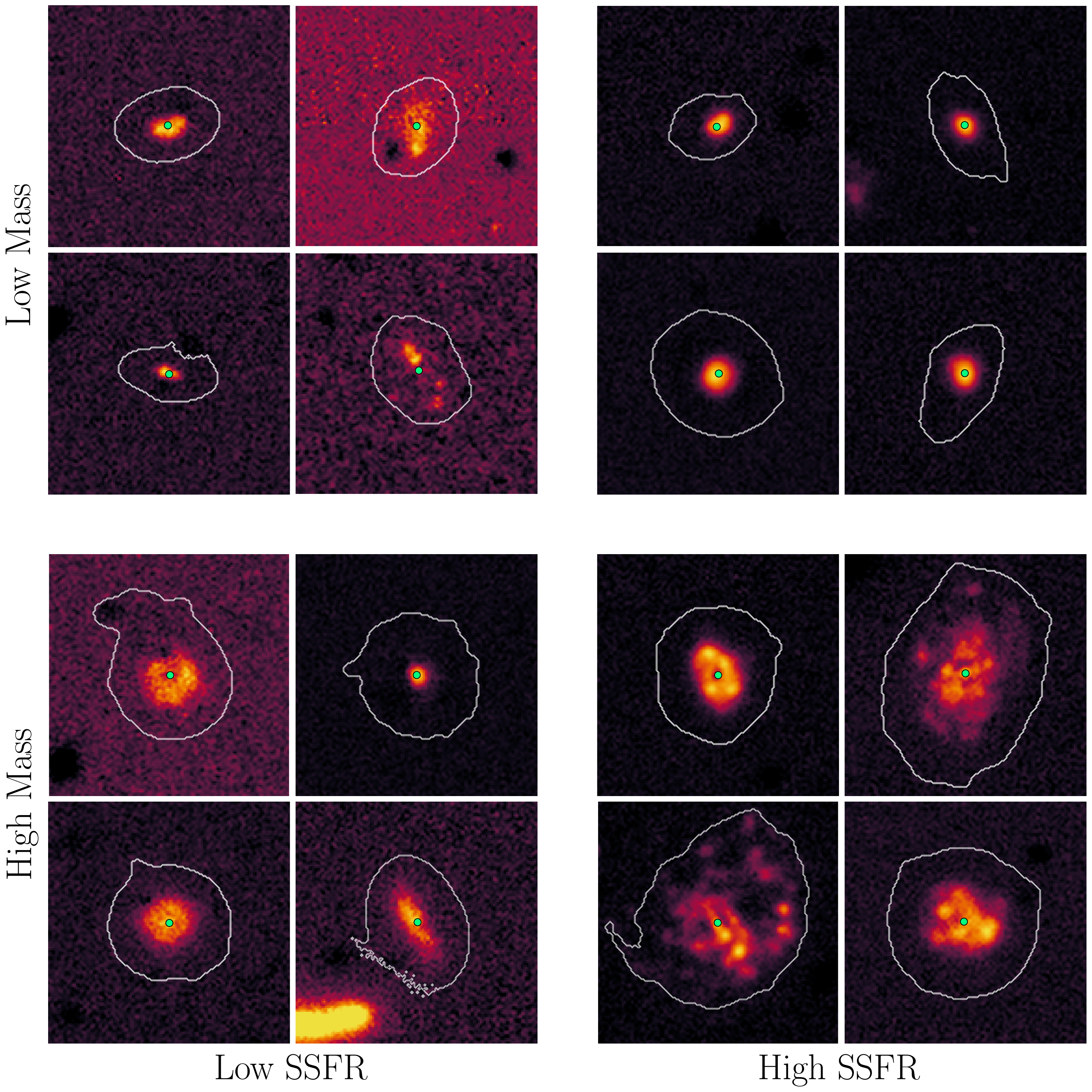}
\caption{The top row shows examples of H$\alpha$ maps for low-mass galaxies with the lowest SSFRs (top left) and highest SSFRs (top right). The white contours indicate the extent of the stellar continuum of the galaxy as measured from the $z$-band images, and the green circles indicate the center of the galaxies, also measured from the $z$-band images. The bottom row shows examples of high-mass galaxies with the lowest SSFRs (bottom left) and the highest SSFRs (bottom right). The actively star-forming low-mass galaxies are all very compact (i.e. have a high Gini coefficient) and largely centralized (low $M_{20}$), with one clump of H$\alpha$ emission near and often slightly offset from the galaxy's center, resulting in high asymmetry. The less active low-mass galaxies are more diffuse and extended. While the high-mass galaxies with high SSFR appear more asymmetric, there is no evident difference in homogeneity of the H$\alpha$ emission distribution between high-mass galaxies with and without vigorous star formation. Complete samples of H$\alpha$ maps for these groups can be found in Appendix~\ref{app:fullgrids}.\label{fig:highlow_mass_ssfr}}
\end{center}
\end{figure*}

    In this section, we present the main results of our morphological analysis, focusing on differences in morphology between the continuum and H$\alpha$ maps, as well as trends with physical properties. In Section~\ref{subsec:comp_ha_cont} we describe the differences in the morphologies of the H$\alpha$ maps and the stellar continua. In Section~\ref{subsec:asymmetry} we present the correlation between asymmetry and SSFR. In Section~\ref{subsec:morph_mass} we describe observed trends with galaxy stellar mass and SSFR. Lastly, in Section~\ref{subsec:morphlowmass} we focus on our results for the lowest-mass galaxies, an especially novel contribution of our sample. 
\subsection{Comparison of H$\alpha$ Maps with Continuum}\label{subsec:comp_ha_cont}

    The Merian medium-band photometry and imaging provide an exciting opportunity to isolate the two-dimensional distribution of H$\alpha$ emission in the observed galaxies and to compare this distribution to the emission of the stellar continuum for a large and consistently processed sample. In Figure~\ref{fig:mass_kdes} we show a comparison of the distribution of morphological parameters measured from the H$\alpha$ maps to those measured from the continuum images (the scaled $z$ band) binned by mass. In every mass bin, the H$\alpha$ maps have significantly higher values of $G$, $M_{20}$, and asymmetry than the continua. The H$\alpha$ maps are overall much less homogeneous and much more asymmetric, with their brightest regions less centrally concentrated. 
    The H$\alpha$ maps also exhibit a larger spread in the distribution of their values of $G$, $M_{20}$, and $A$. Compared to the continua, the H$\alpha$ maps are overall more morphologically diverse and much more irregular. 

    The difference in the distribution of morphological parameters between the H$\alpha$ maps and the continuum images is also evident in \autoref{fig:g_v_m20}, which shows the distribution of the Gini coefficient and $M_{20}$ statistics for the two sets of images, colored by asymmetry. Using broadband morphologies, \citet{Lotz2008} found that merger candidates fall into the upper right quadrant of this parameter space - as defined by the lines shown in the figure. The irregular morphologies of the H$\alpha$ map are sufficiently distinct from their continuum images, and so their $G$-$M_{20}$ distribution cannot be interpreted in this way; the distribution of $G_{{\rm H}\alpha}$ and $M_{20, {\rm H}\alpha}$ covers a larger area of the parameter space than the continuum statistics. Rather than being sensitive to mergers, the H$\alpha$ maps are sensitive to the distribution of star formation surface density.
    
    \subsection{Asymmetry and SSFR}\label{subsec:asymmetry}
    We find a significant correlation between asymmetry and SSFR, a trend that has been previously well established for higher-mass galaxies in both observations and simulations \citep{Yesuf2021, Bottrell2023}. In the top panels of Figure~\ref{fig:sfms_asym}, we show the distribution of SSFR versus stellar mass, colored by asymmetry ($A_{{\rm H}\alpha}$ and $A_{{\rm cont}}$ respectively). In the bottom panels, we show $A_{{\rm H}\alpha}$ and $A_{{\rm cont}}$ as a function of distance from the SFMS ($\Delta$MS) using the fit to our sample as described in Section~\ref{subsec:sfr}. There is a clear and significant correlation between $\Delta$MS and both $A_{{\rm H}\alpha}$ and $A_{{\rm cont}}$, with asymmetry increasing with distance above the MS. This trend holds for both the H$\alpha$ maps and the continuum maps, with the H$\alpha$ maps being more asymmetric overall. 


    \subsection{Morphological Trends with Mass and SSFR}\label{subsec:morph_mass}

    In addition to the unique contribution of our H$\alpha$ distributions, our sample is also novel in its mass range – namely in the large number of galaxies below $\log (M_\star/M_\odot) \lesssim 9.3$. In \autoref{fig:mass_kdes}, we show the distribution of the measured morphological parameters for the H$\alpha$ maps and the continua binned by mass. There is little difference among the distributions for the three highest-mass bins, except for a slight decrease in $M_{20, {\rm cont}}$ and in $G_{{\rm H}\alpha}$ and a slight increase in $A_{{\rm H}\alpha}$ with increasing mass. The values of $G_{{\rm cont}}$, $A_{{\rm cont}}$, and $M_{20, {\rm H}\alpha}$ are otherwise similarly distributed for the higher mass bins. 

    The lowest-mass bin, however, on average has a larger value of $G_{{\rm H}\alpha}$ – i.e. the H$\alpha$ emission of the lowest-mass galaxies is less homogeneously distributed than for the higher-mass galaxies. The lowest-mass galaxies also have lower values of $A_{{\rm H}\alpha}$ on average; they are more symmetric. The main difference between the H$\alpha$ emission of the lowest-mass galaxies and that of the higher-mass galaxies is the lack of diffuse emission and the high concentration (but not necessarily central concentration) of the emission. As discussed in Section~\ref{subsec:quality}, the majority of sources removed from the sample owing to unreliable morphological measurements were in the lowest-mass bin and appear to have diffuse emission. It is possible that the Gini coefficient for the low-mass values could be slightly inflated by this cut, but not sufficiently so as to account for the observed trend.

    We also investigate morphological trends with SSFR as shown in Figure~\ref{fig:ssfr_kdes}. Once again, the increase in asymmetry with SSFR is clear, both in the continuum and in the H$\alpha$ maps. The lack of strong asymmetry trends with mass as discussed above indicates that, despite the correlation between SSFR and stellar mass, asymmetry is independently correlated with SSFR. 

    We see an increase in the Gini coefficient of the H$\alpha$ maps in the highest SSFR bin. We also see a trend of increasing $M_{20}$ for the H$\alpha$ maps, except in the highest SSFR bin, which has a significantly lower median $M_{20}$. There are no trends in these parameters with SSFR for the continuum images. The objects with the highest SSFR tend to be less homogeneous and to have their brightest regions more centrally concentrated. We find that these trends with SSFR are driven primarily by the lowest-mass galaxies, which we investigate more below. Unlike the trends with mass, the quality cuts discussed in Section~\ref{subsec:quality}, which preferentially remove diffuse, low SSFR objects that appear to be more symmetric, would likely lead to the underestimation of the observed trends with SSFR. 
    
    It is possible that the trends with mass discussed above are affected by the mass-dependent H$\alpha$ attenuation discussed in Section~\ref{subsec:morph_dust}. A correction for dust attenuation in the high-mass galaxies would likely result in an increase in the measured values of $G_{{\rm H}\alpha}$ and a decrease in $M_{20, {\rm H}\alpha}$. While we cannot precisely quantify this bias, as described in Section~\ref{subsec:morph_dust} and \autoref{app:dust_morph}, we believe it is below $10\%-20$\%. The mass dependence of dust attenuation could be contributing to the observed trend in $G_{{\rm H}\alpha}$ with mass or masking a trend in $M_{20, {\rm H}\alpha}$ with mass, but we found that applying our derived morphological corrections for dust (see details in \autoref{app:dust_morph}) did not significantly impact the observed trends with SSFR.
    
    \subsection{Morphological Trends with SSFR for the Lowest Mass Bin}\label{subsec:morphlowmass}

    In addition to the aggregate trends with mass and SSFR discussed above, we investigate trends of morphological parameters with SSFR for our galaxies binned by mass. We show the trends in morphological parameters for the H$\alpha$ maps with SSFR for two different mass bins in Figure~\ref{fig:gini_asym}. The bins are selected in order to compare galaxies with distinct physical properties. Galaxies do not begin to form coherent thin disks or spiral arms until at least $\log (M_\star/M_\odot)\gtrsim 9.0$ \citep{Sanchez-Janssen2010, Kado-Fong2020b, Dekel2020} and below this mass are puffier and more spheroidal. We therefore choose our low-mass bin to include all galaxies with $\log (M_\star/M_\odot) \leq 9.3$, which accounts for about 25\% of our total sample. Our higher mass bin includes galaxies with $\log (M_\star/M_\odot)\geq 9.7$, well into the regime of thin disks. We note that while the choice of bins is arbitrary, the trends presented below are not sensitive to small changes in the bin limits. 
    
    The SSFR-asymmetry correlation holds in both the H$\alpha$ maps and the continuum maps for the lowest-mass galaxies, but we also find that the Gini coefficient of the H$\alpha$ maps increases with SSFR for the lowest-mass galaxies while $M_{20}$ decreases. This trend is not evident for the higher mass sources, which show no significant trend in $G_{{\rm H}\alpha}$ or $M_{20, {\rm H}\alpha}$ with SSFR. For the higher mass sources, we do not span a sufficient range in SSFR to conclusively evaluate whether these morphological trends hold for high-mass sources with extremely high SSFR. We note that this class of galaxies is rare locally and it would be challenging to find a large sample with these properties. As discussed in Section~\ref{subsec:quality}, for galaxies in the higher mass bin, the H$\alpha$ emission is likely most strongly attenuated near the center of the galaxies (up to $\sim$\,2-3\,mag)  and minimally affected beyond $\sim$\,1-3\,kpc. While it is possible that our measured morphological parameters are slightly impacted by extinction, we expect this effect to be minimal in our mass and SSFR regime.

    For the lowest-mass galaxies, as SSFR increases, their H$\alpha$ emission becomes both more asymmetric, less homogeneous, and more centralized. Their H$\alpha$ emission – and therefore their sites of recent star formation – is highly compact, and somewhat nuclear but slightly off-center. This trend is not present in the distribution of continuum morphological parameters or in the distributions for the higher-mass galaxies. This result highlights the strength of our sample's unique ability to trace H$\alpha$ emission independently and to probe galaxies at lower masses.     

\section{Discussion}\label{sec:discussion}

In this section, we further contextualize and discuss the implications of our results. First in Section~\ref{subsec:literature} we compare our main morphology trends with results from the literature. In Section~\ref{subsec:asymlowmass} we discuss how the extension of the SSFR--asymmetry trend to lower-mass galaxies should be interpreted. In Section~\ref{subsec:ha_conc} we consider the physical and evolutionary questions raised by the low-mass $G_{{\rm H}\alpha}$--$M_{20, {\rm H} \alpha}$--SSFR trend – namely how does a galaxy end up with such a highly concentrated H$\alpha$ distribution, and what can this tell us about star formation in dwarf galaxies? Lastly, in Section~\ref{subsec:ha_cont_disc} we focus on what new information can be learned from separating the H$\alpha$ emission from the stellar continuum. 

\subsection{Comparison with Morphology Results in the Literature}\label{subsec:literature}
While no previous studies have probed H$\alpha$ morphologies at precisely the same mass and redshift range as our sample, the body of research studying two-dimensional H$\alpha$ emission and the relationship between star formation and morphology is substantial. In this section, we highlight some key results from the literature and demonstrate how our results agree with and extend previously established trends.

One of our main results discussed in Section~\ref{subsec:comp_ha_cont} is the finding that the H$\alpha$ maps in our sample are overall more asymmetric and less homogeneous than the continuum images, with broader distributions of their morphological parameters. A number of other studies –  some comparing broadband images to narrowband H$\alpha$ \citep{Fossati2013, Boselli2015} and some using integral field spectroscopy \citep{Nersesian2023} – have also found H$\alpha$ emission to be significantly more asymmetric than the stellar continuum with larger scatter. \citet{Fossati2013} and \citet{Boselli2015} report mean broadband asymmetries of $\sim$0.1 with a standard deviation of $\sim$0.08 and mean H$\alpha$ asymmetries of $\sim$0.2-0.35 with a standard deviation of $\sim$0.2 in near-perfect agreement with our sample, which has a mean broadband asymmetry of 0.1 with $\sigma= 0.08$ and a mean H$\alpha$ asymmetry of 0.26 with $\sigma=0.2$ (see \autoref{fig:mass_kdes}). We do note that most of these studies are focused on local galaxies, typically with larger masses than we consider in our sample. 

While not all such studies have explicitly considered the Gini coefficient, given their more local samples and therefore higher spatial resolution, some have included the CAS smoothness parameter in their analysis \citep{Boselli2015} and find that H$\alpha$ emission is significantly clumpier than the continuum, again with a broader distribution. This is conceptually similar to our finding for the Gini coefficient, indicating that H$\alpha$ emission is less smooth and homogeneous than the stellar continuum. We note that \citet{Nersesian2023} find a lower value of the Gini coefficient for H$\alpha$ as compared to the continuum, but it is likely that this difference is reflective of our different mass ranges, our different physical resolution, or of our decision to fix the Gini segmentation maps of the H$\alpha$ images to those derived for the continuum images. 

In addition to confirming the general morphological characteristics of the H$\alpha$ maps, previous studies have presented similar findings for trends with mass and SSFR. In particular, a lack of change in asymmetry (both for the H$\alpha$ and for the continuum) with mass has been seen in narrowband surveys \citep{Fossati2013, Boselli2015}. These studies have also found a correlation between asymmetry and SSFR – both for the continuum and for H$\alpha$. Again, while few studies have specifically investigated trends with the Gini coefficient for H$\alpha$, \citet{Boselli2015} did find that the smoothness parameter of the H$\alpha$ maps increases with SSFR; H$\alpha$ emission becomes clumpier as SSFR increases. 

\citet{Yesuf2021} conducted a thorough mutual information analysis and determined that, out of their chosen set of physical parameters, asymmetry is most strongly correlated with $\Delta$MS. Their sample covered a similar redshift range to our sample but did not extend to masses below $\log (M_\star/M_\odot)\approx9.5$. Their morphological parameters were calculated based on broadband imaging and not H$\alpha$ imaging. \citet{Yesuf2021} proposed that the asymmetry-$\Delta$MS relation can perhaps be explained by asymmetric cold accretion leading to elevated star formation or by structural asymmetries induced by interactions or minor mergers.

\citet{Bottrell2023} further investigated the physical explanations for the observationally established relationship between asymmetry and high SSFR using simulated HSC-SSP imaging from IllustriusTNG. They created synthetic $i$-band imaging for galaxies at redshifts similar to and higher than our sample, with typical masses higher than what we cover. They were able to reproduce the asymmetry-$\Delta$MS trend with their simulated galaxies and argue that mini mergers are the dominant source of enhanced asymmetry and SSFR. They find that mini mergers are more common and lead to smaller but more prolonged boosts in SSFR and asymmetry than minor or major mergers. 

Overall, our main findings agree with existing results surrounding nonparametric morphology of H$\alpha$ emission as compared to continuum emission. They also generally agree with morphological trends with mass, SSFR, and $\Delta$MS, particularly trends in asymmetry. Our sample extends to lower masses than previous studies at comparable redshifts. We consider the implications of this new mass bin below.

\subsection{Asymmetry as a High-SSFR Predictor at Low Mass}\label{subsec:asymlowmass}

The correlation between asymmetry and SSFR (or between asymmetry and $\Delta$MS) has been previously established both in observations and in simulations, as described above \citep{Fossati2013, Boselli2015, Yesuf2021, Bottrell2023}. The physical explanations for this relationship focus on two main phenomena: galaxy-galaxy interactions and cold gas accretion. But what is the relative importance of each of these processes, and how do they change at low mass?

It is largely agreed upon that mergers lead to morphological asymmetry, with the lower mass companion being more strongly affected by the interaction if it survives \citep{DePropris2007, Casteels2014}. Mergers are also known to lead to strong enhancements in star formation \citep{Keel1985, Ellison2013, Kaviraj2014}. However, major mergers occur relatively infrequently \citep{Lin2004, Patton2008, LopezSanjuan2009}, and certainly not frequently enough to explain the ubiquity of the SSFR-asymmetry trend. \citet{Bottrell2023} found that major and minor mergers were far too rare to cause observed asymmetry and SFR enhancements, but mini mergers (mergers with companions with greater than a 1:10 ratio of stellar masses) occurred sufficiently frequently. They also found that a more dramatic mass ratio in a merger led to a longer-lived increase in asymmetry and SFR. Overall, they concluded that in their simulations mini mergers accounted for twice as much asymmetry as minor and major mergers combined. 

Can this same explanation be applied to dwarfs? The merger-asymmetry boost has been shown to hold at lower masses \citep{Starkenburg2016b, Guzman-Ortega2023}, but \citet{Martin2021} showed that mergers can only account for a small amount of the observed morphological disturbances in dwarf galaxies. Rather, they argue that nonmerger interactions are more important in shaping the morphologies of dwarf galaxies. \citet{Besla2018} found that while the major merger rate decreases with decreasing stellar mass for dwarf galaxies, the fraction of dwarfs with a close companion of comparable mass increases with decreasing mass. Hence, for dwarfs, major mergers are likely not a dominant process in creating asymmetry. Instead, less dramatic but more common interactions with nearby galaxies could play a significant role. Similarly, the connection between mergers/interactions and enhanced star formation has been shown to hold for dwarf galaxies \citep{Stierwalt2015, Starkenburg2016a, Kado-Fong2020a, Kado-Fong2023}. It therefore seems reasonable that the same class of interaction-related explanations should hold for the asymmetry-SSFR trend at low mass.

The second important physical phenomenon invoked to explain the asymmetry-SSFR correlation is cold accretion. Higher-mass galaxies, which have lower gas fractions than dwarf galaxies, require additional material to fuel sustained star formation. Galaxies are thought to obtain this extra gas by cold accretion along dark matter filaments \citep{Sancisi2008, Sanchez-Almeida2014}, which has been shown to lead to asymmetries in galaxy morphologies and is well associated with an increase in SSFR. 

It is important to note that the accretion explanation is not completely independent of the merger explanation; it is challenging to determine exactly how many mini mergers occur independently of cold accretion along filaments. Low-mass galaxies could be accreted along preexisting streams of material, as opposed to being independently gravitationally attracted to merge with the primary galaxy. This distinction is especially unclear for dwarf galaxies, for whom mini mergers would involve systems of very low mass, possibly even individual clumps of cold gas with no stars at all. 

Additionally, because dwarf galaxies are gas-rich \citep{Bradford2015}, the addition of fuel \citep{Hallenbeck2012, Richtler2018, Ju2022} may be less important in dwarfs than any accompanying dynamical disturbance.  It is likely that the accretion-enhanced asymmetry and SSFR explanation for dwarf galaxies is less conceptually distinct from the impact of mergers.

\subsection{High SSFR Requires Concentrated H$\alpha$ for Low-mass Galaxies}\label{subsec:ha_conc}

One of the most striking results of our study is that the H$\alpha$ emission occurs in a small number of marginally resolved regions in the high-SSFR, low-mass galaxies. (No such trend is observed in the higher-mass galaxies, but we note that our sample does not probe the same SSFR regime for these sources.) Few if any of the rigorously star-forming low-mass galaxies have anything resembling a diffuse H$\alpha$ map as demonstrated quantitatively in Figure~\ref{fig:gini_asym} and shown qualitatively with the maps themselves in \autoref{fig:highlow_mass_ssfr} (an extension of \autoref{fig:highlow_mass_ssfr} with complete samples of H$\alpha$ maps for the groups shown is included in Appendix~\ref{app:fullgrids}). What can this trend tell us about bursting dwarf galaxies and the processes that lead to dramatic star formation enhancements? 

We should first note that while H$\alpha$ emission does not explicitly trace the presence of molecular gas, such material is a necessary prerequisite for recent star formation. Though in dwarf galaxies it is very challenging to detect CO directly \citep{Leroy2005}, recent high-resolution studies of higher-mass galaxies have shown that on the scale of our physical resolution ($\gtrsim$1 kpc) the distribution of CO is very similar to the H$\alpha$ emission \citep{Pan2022}. Hence, we can reasonably conclude that there is dense molecular gas at the sites of highly concentrated H$\alpha$ emission. 

Dwarf galaxies differ fundamentally from larger galaxies, not only in their aggregate mass but also in their structure. While galaxies with $\log (M_\star/M_\odot) > 9.3$ have coherent large-scale structures including thin disks, spiral arms, or galactic bars, dwarf galaxies are typically puffy with shallow potentials and no such coherent large-scale features \citep{Sanchez-Janssen2010, Kado-Fong2020b, Dekel2020}. These structural distinctions help explain observed differences in composition and in star formation behavior between dwarfs and high-mass galaxies. Under the pressure-regulated, feedback-modulated (PRFM) model of star formation \citep{Ostriker2022}, the rate at which a galaxy forms stars is set by its midplane pressure. Due to their lower mass surface densities and larger scale heights, dwarfs tend to have lower star formation efficiencies and so form fewer stars and retain a larger fraction of their gas. 

Given these differences, is it surprising that we do not see evenly distributed H$\alpha$ emission – i.e. evenly spatially distributed star formation and molecular gas –  in the most active low-mass galaxies? As discussed above, enhanced star formation is most often thought to result from interactions with other galaxies and/or from accretion of cold gas. For a typical Milky Way-like galaxy, a mini merger or an interaction may cause enhanced star formation via compression in molecular clouds in the disk but is unlikely to cause a galaxy-scale instability - leading to spatially distributed star formation. (We note that in massive galaxies with higher SSFRs than what we probe in our sample, highly concentrated central star formation has been observed – as in ultraluminous infrared galaxies.) Conversely, for a low-mass galaxy without any such stabilizing structure and a much shallower potential, the same interaction could destabilize the disk, causing the galaxy's gas reservoir to collapse toward its center. This type of rapid compression would result in highly concentrated molecular gas and star formation. It is therefore unlikely that we would see a low-mass galaxy with extremely active star formation evenly distributed over the galaxy. 

It is also unlikely that a galaxy in this mass range could sustain such a high SSFR concentrated in a small central area if it were in dynamical equilibrium because the implied gas surface densities would be much higher than expected in dwarfs. Specifically, given their implied star formation rate surface densities, the extended Kennicutt-Schmidt relation \citep[see, e.g.,][]{delosreyes2019} implies extremely high molecular gas fractions ($>\! 0.9$) for these galaxies; such high molecular gas surface densities significantly exceed the surface densities expected in low-mass galaxies \citep[see, e.g.,][]{Leroy2008, delosreyes2019, Kado-Fong2022}. We elaborate on this estimate and sketch out a toy model in Appendix~\ref{s:appendix:prfm} to demonstrate that it is highly unlikely that these sources are in dynamical equilibrium and to further support our interpretation of an external nonequilibrium process for starbursts in the low-mass sample.

The prevalence of a near-central clump of H$\alpha$ emission in the lowest-mass galaxies fits well within the larger picture of star formation in dwarf galaxies. In addition to their lower star formation rate surface densities, dwarfs have also been shown to have significantly burstier star formation histories, forming large fractions of their stellar populations in discrete episodes of star formation rather than forming stars at a constant rate for a prolonged period of time. In fact, the burstiness of a galaxy's star formation history is believed to be a function of its stellar mass \citep{Weisz2012, Guo2016, Sparre2017, Emami2019, Atek2022,Pan2023}. Starbursting dwarf galaxies are typically dominated by a small number of large \ion{H}{2} regions \citep{Lee2009a, Zick2018}, sometimes even a single \ion{H}{2} region for the lowest-mass galaxies. While our imaging cannot resolve individual \ion{H}{2} regions, it is likely that the high heterogeneity of the H$\alpha$ maps for the low-mass, high-SSFR galaxies is driven by this phenomenon. 

This also helps explain why we do not see the $G_{{\rm H}\alpha}$-SSFR or $M_{20,{\rm H}\alpha}$-SSFR trends for higher mass. They tend to be less bursty and more efficient in their star formation, with sites of star formation more spatially distributed. We note that our sample does not include any high-mass objects with extremely high SSFRs, so we are unable to comment on the morphology of this class of rare and extreme objects.

\subsection{What Can Be learned from Separating H$\alpha$ Emission from the Stellar Light?}\label{subsec:ha_cont_disc}

Our finding that nearly all of the low-mass, high-SSFR galaxies in our sample have H$\alpha$ distributions dominated by a small number of marginally resolved clumps can also shed light on questions about the physical interpretations of existing dwarf galaxy classifications. For decades, there has been much interest in identifying, classifying, and understanding blue compact dwarfs \citep[BCDs, e.g.,][]{Thuan1981,GildePaz2003}, which, as their name implies, are characterized by their blue color and compact light distribution. Do BCDs constitute an independent population of objects, or do they represent a transient stage in the life cycle of a typical dwarf galaxy? How do these galaxies fit into the larger narrative of dwarf galaxy evolution? Significant evidence exists that BCDs are triggered by interactions with their environment \citep[e.g.,][]{Taylor1997, Bekki2008, Ashley2017} – interactions that are similar to the proposed triggering scenarios described above. Our results provide further evidence for these theories, supporting the interpretation that BCDs represent a transient stage in a dwarf galaxy's life cycle – one of active star formation that is triggered by interactions with its environment.

Dwarf galaxies are bursty, and our analysis shows that bursting dwarf galaxies are often if not always highly concentrated, with a small number of compact star formation sites near, but slightly offset from, the center of the galaxy. Nonbursting low-mass galaxies – those with lower SSFRs – are less compact and more homogeneous. It is reasonable to conclude that most typical dwarf galaxies go through stages where they are blue and compact (the same stages where they are highly star-forming) and then settle down to a more diffuse morphology. Dwarf classifications based on compactness – especially those using broadband imaging that contains emission representative of star formation – may therefore fundamentally represent cuts based on SSFR. 

\section{Summary}\label{sec:summary}
We have presented an analysis of the relationship between nonparametric morphological parameters and physical properties of galaxies from the first data release of the optical medium-band Merian survey. We included only galaxies with spectroscopic redshift measurements at $0.064<z<0.1$ and with $\log (M_\star/M_\odot) < 10.3$ – a total of \specsamplesizeinbandlowmassallphot\ sources.

We developed a new method for estimating and subtracting the continuum through the Merian N708 filter using a lookup table of synthetic spectra generated using SED fitting results from low-mass GAMA galaxies. We used this approach to make maps of the H$\alpha$ emission and to measure the galaxies' star formation rates. We adapted the publicly available \texttt{Python} package \texttt{statmorph} \citep{Rodriguez-Gomez2019} to measure nonparametric morphological parameters ($A$, $G$, $M_{20}$) for both the continuum images (the scaled $z$-band images) and the H$\alpha$ maps. We ultimately obtained reliable measurements for \postmorphsample\ sources. 

We investigated how these parameters differ between the continuum and the H$\alpha$ emission and how they vary with physical properties. Our main findings are as follows:

\begin{itemize}
    \item The morphology of the H$\alpha$ maps varies substantially from the morphology of the continua for galaxies of all masses and star formation rates. The H$\alpha$ maps are generally more asymmetric, with higher values of the Gini coefficient and $M_{20}$. The dramatic differences in the morphology of the H$\alpha$ emission and the continuum highlight the value of isolating the H$\alpha$ emission and morphology for probing environmental links with star formation. 
    \item Asymmetry – of both the continuum and H$\alpha$ emission – increases with distance above the SFMS. The galaxies with the highest values of SSFR also have among the highest values of asymmetry. We confirm the existence of this previously reported trend, extending it to lower-mass galaxies and showing that it also holds for H$\alpha$ emission. 
    \item We find no significant trends with mass in asymmetry for the continuum emission, but we find that the H$\alpha$ asymmetry increases slightly with mass. While there is similarly no significant trend for $G$ or $M_{20}$ measured from the continuum images, we do find that the $G_{{\rm H}\alpha}$ decreases slightly with increasing mass. The lowest-mass galaxies have H$\alpha$ emission that is less homogeneous and more symmetric than higher-mass galaxies. 
    \item We find that the Gini coefficient of the H$\alpha$ emission is significantly higher for the galaxies with the highest SSFR and that the $M_{20}$ statistic is lower. We investigate these trends for different mass bins and find that while the highest-mass galaxies show little change in $G_{{\rm H}\alpha}$ or $M_{20,{\rm H}\alpha}$ with increasing SSFR, the $G_{{\rm H}\alpha}$ for the lowest-mass galaxies increases significantly and $M_{20,{\rm H}\alpha}$ decreases.
\end{itemize}

Our finding that low-mass, highly star-forming galaxies have strikingly different morphologies than their higher mass counterparts provides insight into the mechanisms that lead to starbursts in dwarfs. Dwarfs are gas-rich and without stable large-scale structures in their disks. They are more sensitive to the dynamical effects of interactions and as such can more easily be dramatically disturbed. The ubiquity of the concentration of H$\alpha$ emission in these systems to a small number of marginally resolved clumps near but slightly offset from the galaxy's center indicates that interactions (leading to collapse of molecular gas) are likely the dominant pathway for triggering a starburst in dwarfs. 
\\
\\
A.M. acknowledges support from the National Science Foundation Graduate Research Fellowship under Grant No. 2039656.
S.H. thanks the support from National Natural Science Foundation of China (No. 12273015). J.G. and A.L. are supported by a National Science Foundation Astronomy and Astrophysics Research Grant under Grant Nos. 1007052 and 2106839 respectively.

The Hyper Suprime-Cam (HSC) collaboration includes the astronomical communities of Japan and Taiwan and Princeton University. The HSC instrumentation and software were developed by the National Astronomical Observatory of Japan (NAOJ), Kavli Institute for the Physics and Mathematics of the Universe (Kavli IPMU), University of Tokyo, High Energy Accelerator Research Organization (KEK), Academia Sinica Institute for Astronomy and Astrophysics in Taiwan (ASIAA), and Princeton University. Funding was contributed by the FIRST program from the Japanese Cabinet Office, Ministry of Education, Culture, Sports, Science and Technology (MEXT), Japan Society for the Promotion of Science (JSPS), Japan Science and Technology Agency (JST), Toray Science Foundation, NAOJ, Kavli IPMU, KEK, ASIAA, and Princeton University.

This project used data obtained with the Dark Energy Camera (DECam), which was constructed by the Dark Energy Survey (DES) collaboration. Funding for the DES Projects has been provided by the US Department of Energy, the US National Science Foundation, the Ministry of Science and Education of Spain, the Science and Technology Facilities Council of the United Kingdom, the Higher Education Funding Council for England, the National Center for Supercomputing Applications at the University of Illinois at Urbana-Champaign, the Kavli Institute for Cosmological Physics at the University of Chicago, Center for Cosmology and Astro-Particle Physics at the Ohio State University, the Mitchell Institute for Fundamental Physics and Astronomy at Texas A\&M University, Financiadora de Estudos e Projetos, Fundação Carlos Chagas Filho de Amparo à Pesquisa do Estado do Rio de Janeiro, Conselho Nacional de Desenvolvimento Científico e Tecnológico and the Ministério da Ciência, Tecnologia e Inovação, the Deutsche Forschungsgemeinschaft and the Collaborating Institutions in the Dark Energy Survey.

The Collaborating Institutions are Argonne National Laboratory, the University of California at Santa Cruz, the University of Cambridge, Centro de Investigaciones Enérgeticas, Medioambientales y Tecnológicas–Madrid, the University of Chicago, University College London, the DES-Brazil Consortium, the University of Edinburgh, the Eidgenössische Technische Hochschule (ETH) Zürich, Fermi National Accelerator Laboratory, the University of Illinois at Urbana-Champaign, the Institut de Ciències de l’Espai (IEEC/CSIC), the Institut de Física d’Altes Energies, Lawrence Berkeley National Laboratory, the Ludwig-Maximilians Universität München and the associated Excellence Cluster Universe, the University of Michigan, NSF’s NOIRLab, the University of Nottingham, the Ohio State University, the OzDES Membership Consortium, the University of Pennsylvania, the University of Portsmouth, SLAC National Accelerator Laboratory, Stanford University, the University of Sussex, and Texas A\&M University.

The authors are pleased to acknowledge that the work reported on in this paper was substantially performed using the Princeton Research Computing resources at Princeton University which is consortium of groups led by the Princeton Institute for Computational Science and Engineering (PICSciE) and Office of Information Technology's Research Computing.

\textit{Software}: \texttt{NumPy} \citep[\url{https://numpy.org},][]{Harris2020}, \texttt{Astropy}  \citep[\url{https://astropy.org},][]{Astropy2013, Astropy2018, Astropy2022}, \texttt{Matplotlib} \citep[\url{https://matplotlib.org},][]{Hunter2007}, \texttt{SciPy} \citep[\url{https://scipy.org},][]{Virtanen2020}, \texttt{FSPS} \citep[\url{https://github.com/cconroy20/fsps},][]{Conroy2009, Conroy2010}, \texttt{Python-FSPS} \citep[\url{https://dfm.io/python-fsps/current/},][]{Johnson2021}, \texttt{statmorph}  \citep[\url{https://statmorph.readthedocs.io/en/latest/},][]{Harris2020}, \texttt{SEP}  \citep[\url{https://sep.readthedocs.io/en/v1.1.x/},][]{Bertin1996, Barbary2016}.

\appendix

\section{Correction for [\ion{N}{2}] and [\ion{S}{2}] emission} \label{app:emission}

To isolate the H$\alpha$ emission, we need to correct for weaker emission lines that fall within the N708 filter. In particular, this includes [\ion{N}{2}] ($\lambda\lambda$ 6548, 6584) and [\ion{S}{2}] ($\lambda\lambda$ 6717, 6732). To estimate the necessary correction as a function of redshift, we calculate the EW of the SDSS spectra integrated over the wavelength range of the entire N708 filter (to emulate the value of H$\alpha$+\ion{N}{2}+\ion{S}{2} that we measure through the medium-band filter) and compare to the integrated spectroscopic value of H$\alpha$ EW. We divide our sample into three mass bins and parameterize the correction as a function of redshift as
\begin{equation}
c(z) = 
\left\{
    \begin{array}{lr}
        c_1 & \text{if } z \leq 0.074 \\
        \frac{c_2 - c_1}{0.09}(z-0.074) + c_1 & \text{if } 0.074 < z < 0.083 \\
        c_2 & \text{if } 0.083 \geq z
    \end{array}
\right.
\end{equation}
where
\begin{equation}
\left\{
    \begin{array}{lr}
        c_1= 1.39,c_2= 1.22& \text{if } \log (M_\star/M_\odot) \leq 9.2 \\
        c_1= 1.77,c_2= 1.39& \text{if } 9.2 < \log (M_\star/M_\odot) < 9.8 \\
        c_1= 1.85,c_2= 1.48& \text{if } 9.8 \geq \log (M_\star/M_\odot)
    \end{array}
    \right.
\end{equation}

We note that this correction is based solely on the SDSS spectra, which does not include the lowest-mass sources in our catalog. Given that the correction for contaminating lines is smaller for lower masses and the SDSS subsample does contain a number of sources with $\log(M_\star/M_\odot < 9)$, the empirically derived correction should still apply for the lowest-mass sources. As shown in \autoref{fig:catalogew}, the H$\alpha$ EWs we measure from our photometry agree well with the spectroscopic values, regardless of the source of the spectra.

\section{Correcting $L_{{\rm H}\alpha}$ for dust extinction}\label{app:dust} 

The extinction of H$\alpha$ emission is often computed via the Balmer decrement, which compares the observed relative strengths of the H$\alpha$ and H$\beta$ lines to what is theoretically expected without extinction. Though H$\alpha$ and H$\beta$ fluxes are available for a subset of our sample, we found the scatter in the extinction estimates derived from the Balmer decrement to be exceedingly large for the GAMA sources. Using these values to correct for extinction would at least double the uncertainty in our estimates of the SFR.

In order to apply a consistent correction to all of our sources and to avoid unduly increasing the scatter and uncertainty in our measured SFRs, we choose not to use the Balmer lines to correct for extinction. Instead, we use the SED-fitted values of $E(B-V)$ for the entire GAMA DR3 catalog \citep{Taylor2011} to derive an empirical relation between $A({{\rm H}\alpha})$ and stellar mass. (We note that elsewhere we use $A_{{\rm H}\alpha}$ to refer to the asymmetry of the H$\alpha$ emission; in this section $A({{\rm H}\alpha})$ refers to attenuation.) We use a \citet{Calzetti2000} attenuation law to compute the extinction at 6563\,\AA\ and double this value to account for the increased dust extinction in \ion{H}{2} regions around young stars \citep{Calzetti1994,Charlot2000}. We fit a quadratic function to the GAMA sources with $\log (M_\star / M_\odot)>9$ and use a constant value for sources with lower mass, similar to the approach in \citet{Lee2009b}. Such low-mass sources are typically metal poor and have minimal extinction, which varies weakly if at all with mass.

Our final correction for internal extinction is calculated as
\begin{equation}
A({{\rm H}\alpha}) = 
\left\{
    \begin{array}{lr}
        0.44 & \text{if } \log(M_\star/M_\odot) < 9 \\
         0.16 - 2.8 \cdot\log(M_\star/M_\odot) + 13.1\cdot\left[\log(M_\star/M_\odot)\right]^2 & \text{if } \log(M_\star/M_\odot) > 9 
    \end{array}
\right.
\end{equation}

We also correct for galactic extinction using the SFD dust maps \citep{Schlegel1998, Schlafly2011, Green2018}. The median correction for foreground extinction is 0.1 mag.

\section{{Estimating the impact of dust attenuation on measured morphological parameters}}\label{app:dust_morph}

In addition to correcting the integrated H$\alpha$ luminosities for extinction, it is necessary to consider the impact of nonuniformly distributed dust on the values of our measured morphological parameters. Previous work has shown that dust attenuation typically follows a radial gradient, with the strongest attenuation concentrated in the galaxy's center \citep{Boissier2004, Boissier2007, Nelson2016, Kahre2018, Jafariyazani2019}. As discussed in Section~\ref{subsec:morph_dust}, we are confident that this effect is negligible for galaxies with stellar masses below $\log(M_\star/M_\odot)<9.2$ and minimal for $\log(M_\star/M_\odot)<9.8$. A true correction for two-dimensional attenuation would require resolved dust maps or measurements of the Balmer decrement, but to estimate upper limits on the potential biasing effect of dust on $G_{{\rm H}\alpha}$, $M_{20, {\rm H}\alpha}$, and Asym$_{{\rm H}\alpha}$ (notated differently in this section to distinguish from the attenuation $A({{\rm H}\alpha})$) for higher-mass galaxies, we conduct a simplified experiment based on the results of \citet{Nelson2016}. 
Using spatially resolved maps of H$\alpha$ and H$\beta$ for over 600 galaxies at $z\approx1.4$, \citet{Nelson2016} demonstrated the radial dependence of $A({{\rm H}\alpha})$ and presented a parameterization as a function of stellar mass and distance from the galaxy's center: 

$$A({\rm{H}\alpha}) = b + c \log(r)$$
where $r$ is the distance from the galaxy center in kpc and $b$ and $c$ are empirically derived functions of the stellar mass:
$$b = 0.9 + [\log(M_\star/M_\odot) - 10]$$
$$c = -1.9 -2.2\cdot[\log(M_\star/M_\odot) - 10].$$

For each galaxy in our sample, we create a map of $A({\rm{H}\alpha})$ using the asymmetry center derived from the continuum image (as described in Section~\ref{subsec:morph_params}) to calculate $r$. We smooth the $A({\rm{H}\alpha})$ map to the PSF of the N708 image and adjust the flux as $F_{\rm adjusted} = F \cdot 10^{A({\rm{H}\alpha})/2.5}$ for each pixel. Using these ``dust-corrected" maps, we recompute the H$\alpha$ morphological parameters as described in Section~\ref{subsec:morph_params}. We find that for galaxies with $\log(M_\star/M_\odot)>9.8$, on average $G_{{\rm H}\alpha}$ increases by $\sim$12\%, $M_{20, {\rm H}\alpha}$ decreases by $\sim$28\%, and Asym$_{{\rm H}\alpha}$ increases by $\sim$4\%. We find that the increase in $G_{{\rm H}\alpha}$ and decrease in $M_{20, {\rm H}\alpha}$ grow as a function of stellar mass, which is likely reflective of the fact that the attenuation correction itself is an explicit function of mass. There is no evident trend in Asym$_{{\rm H}\alpha}$ with mass, and the scatter in the fractional change is $\sim$13\%, so we do not believe that the asymmetry is systematically biased by dust.

We also caution that the biases reported above in $G_{{\rm H}\alpha}$ and $M_{20, {\rm H}\alpha}$ represent high upper limits and are in fact certainly overestimates. The galaxies included in \citet{Nelson2016} are at much higher redshift than those in our sample and have SFRs that are nearly 10 times higher than our sources at comparable mass. They are therefore significantly dustier than our galaxies, and the estimates for $A({\rm{H}\alpha})$ are surely higher than the true values for these sources. But perhaps the most important complication is that in reality the distribution of dust and the impact of attenuation are much more complex than a simple radial profile. In particular, \ion{H}{2} regions (and therefore bright clumps of H$\alpha$ emission) are more strongly attenuated (by a factor of 2-3) than other regions of the galaxy. Correcting for this effect in addition to the radial dependence would diminish the impact of the correction on $M_{20, {\rm H}\alpha}$ as the dust correction would not be purely centralized. Increasing the flux preferentially in the center of the galaxy leads to a smaller value of $M_{20, {\rm H}\alpha}$ essentially by definition - the brightest regions of the map are more centrally concentrated. The impact of a clumpy dust correction on Asym$_{{\rm H}\alpha}$ or $G_{{\rm H}\alpha}$ is less clear and, in practice, is certainly dependent on the morphology of a particular galaxy.\\

\section{Full samples of H$\alpha$ maps with highest and lowest mass and SSFR} \label{app:fullgrids}

\begin{figure*}[t]
\begin{center}
\includegraphics[width = 0.9\linewidth,angle=0]{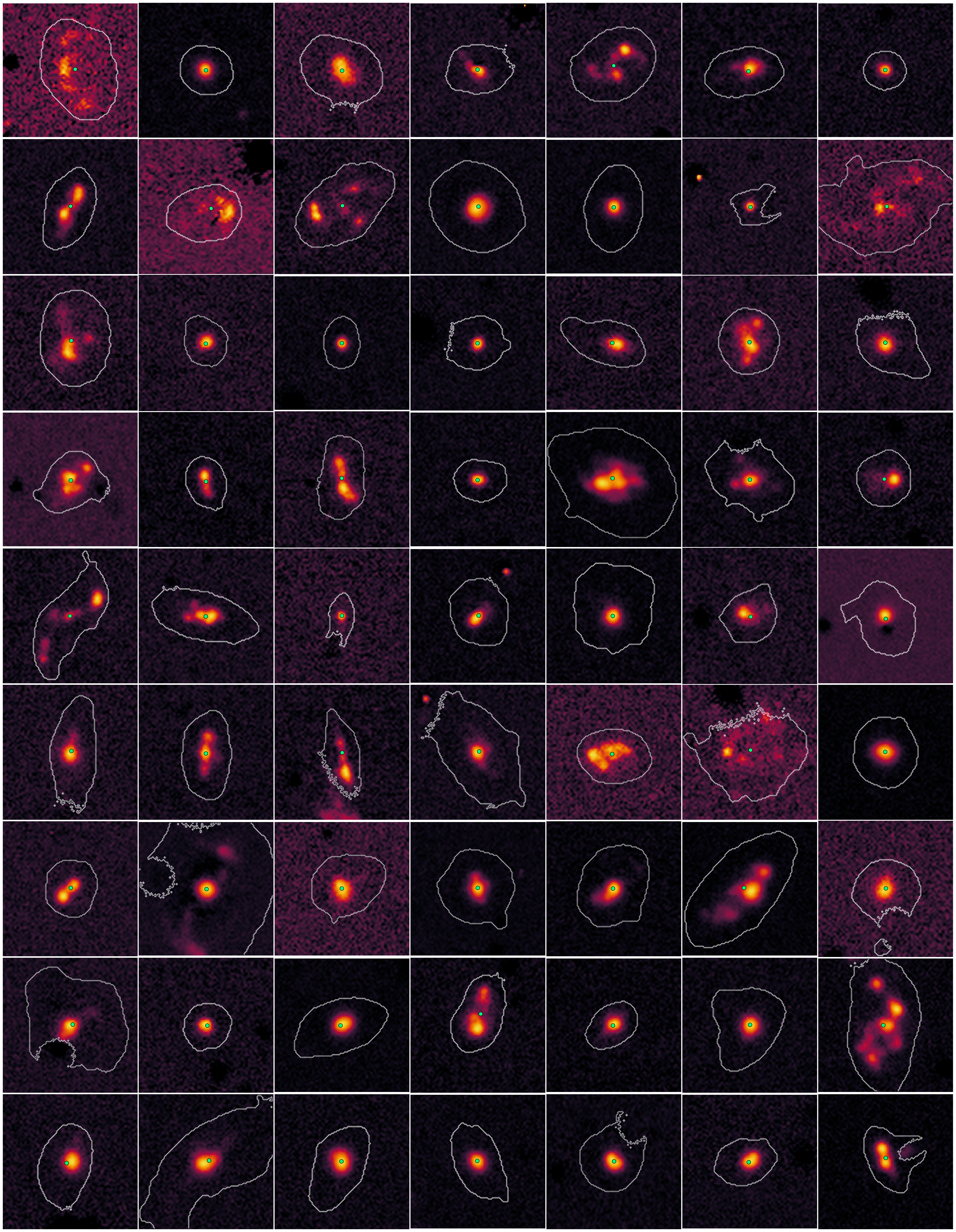}
\caption{H$\alpha$ maps for the galaxies in the lowest-mass bin ($\log (M_\star/M_\odot) <9.3)$ with the highest SSFR (top SSFR quartile of low-mass sources). The white contours show the segmentation map derived from the continuum images and used to calculate the galaxies' global H$\alpha$ flux and star formation rate. The green circles indicate the center of the galaxy as measured from the continuum images. The H$\alpha$ emission is generally very compact and centralized. \label{fig:lowmasshighssfr}}
\end{center}
\end{figure*}

\begin{figure*}[t]
\begin{center}
\includegraphics[width = 0.9\linewidth,angle=0]{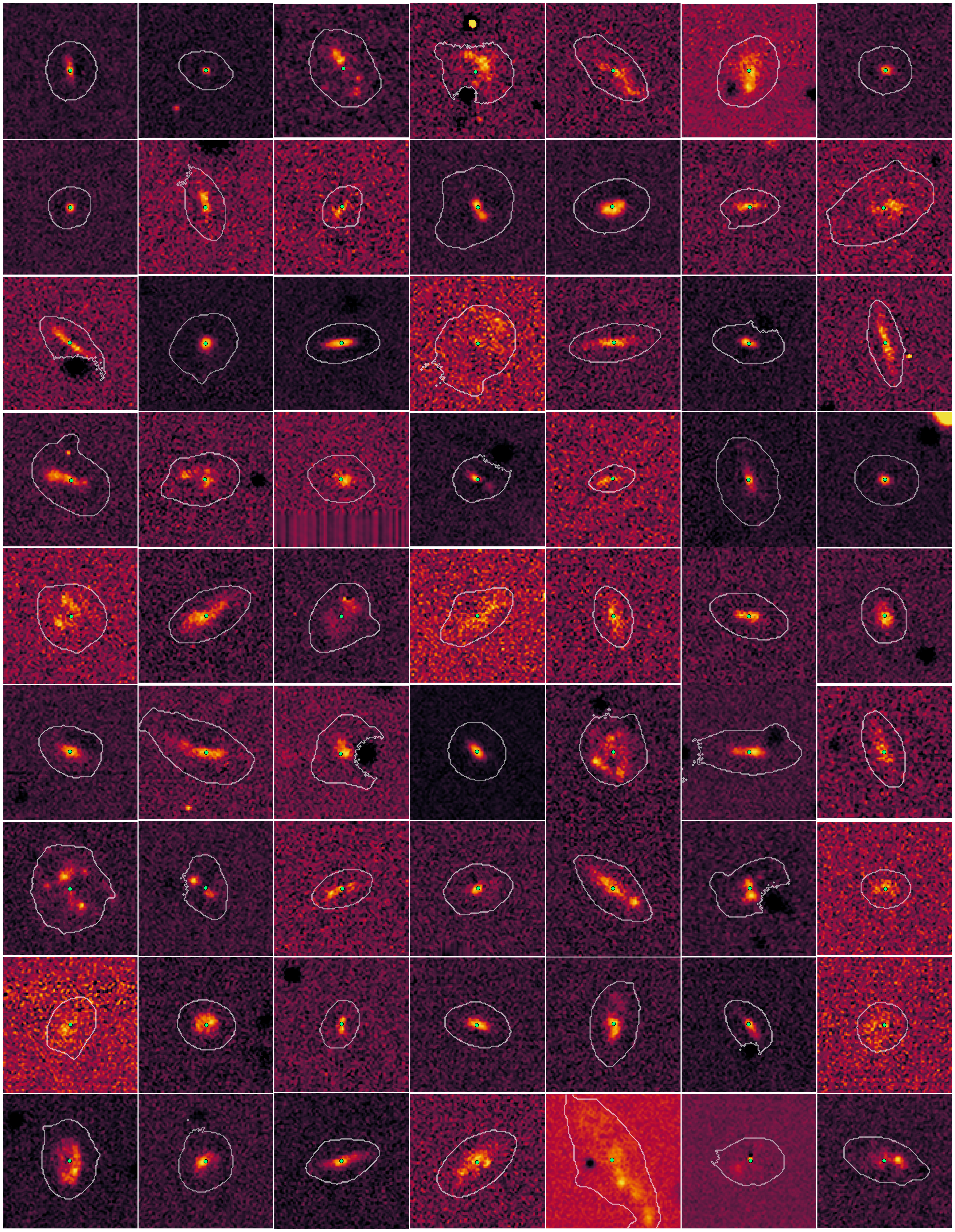}
\caption{Same as \autoref{fig:lowmasshighssfr} but for the galaxies in the lowest-mass bin ($\log (M_\star/M_\odot) <9.3)$ with the lowest SSFR (bottom SSFR quartile of low-mass sources). The H$\alpha$ emission is generally more diffuse and more symmetric than that of the sources shown in \autoref{fig:lowmasshighssfr}.  \label{fig:lowmasslowssfr} }
\end{center}
\end{figure*}

\begin{figure*}[t]
\begin{center}
\includegraphics[width = 0.9\linewidth,angle=0]{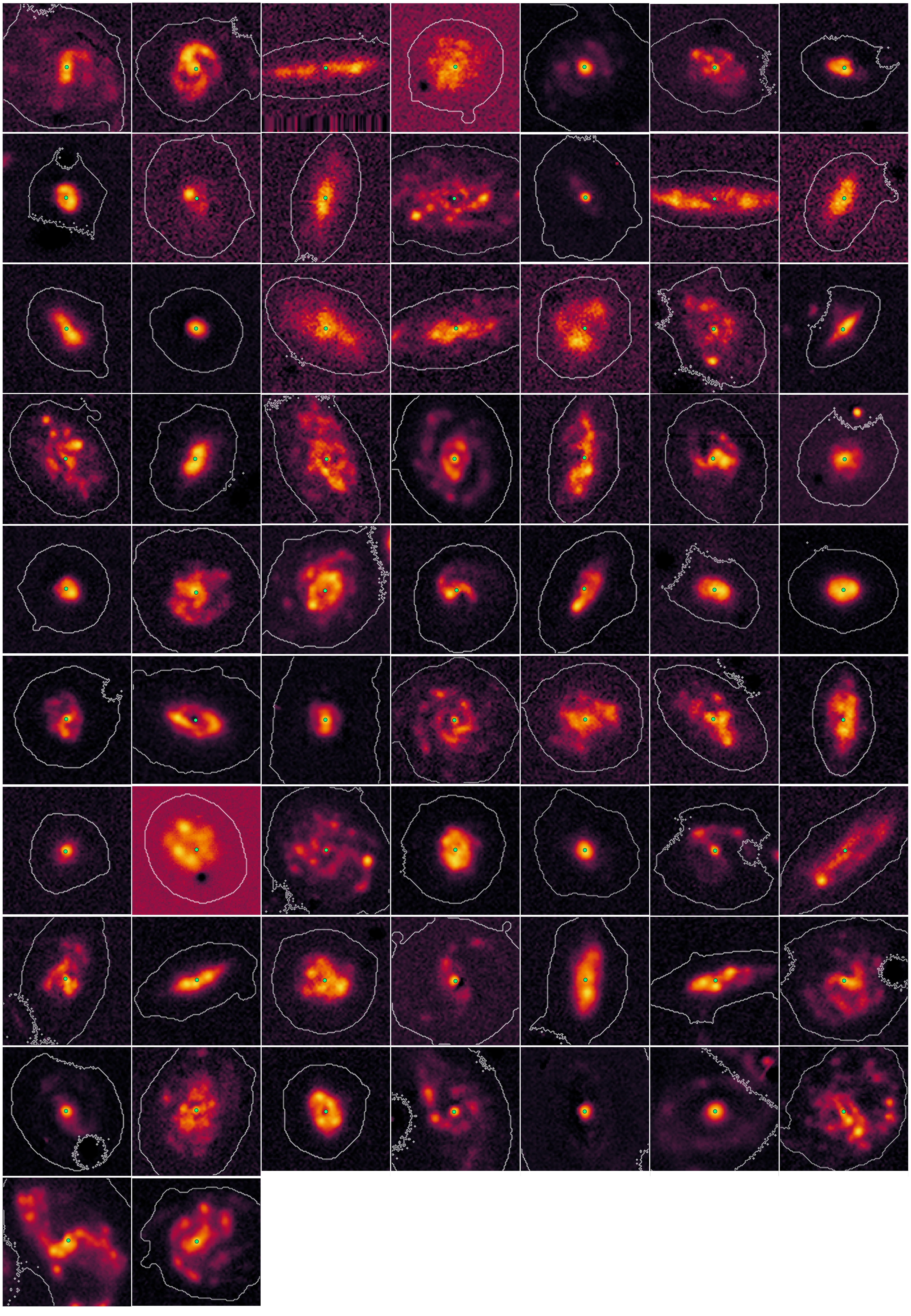}
\caption{Same as \autoref{fig:lowmasshighssfr} but for the galaxies in the highest-mass bin ($\log (M_\star/M_\odot) >10.1)$ with the highest SSFR (top SSFR quartile of high-mass sources). \label{fig:highmasshighssfr} }
\end{center}
\end{figure*}

\begin{figure*}[t]
\begin{center}
\includegraphics[width = 0.9\linewidth,angle=0]{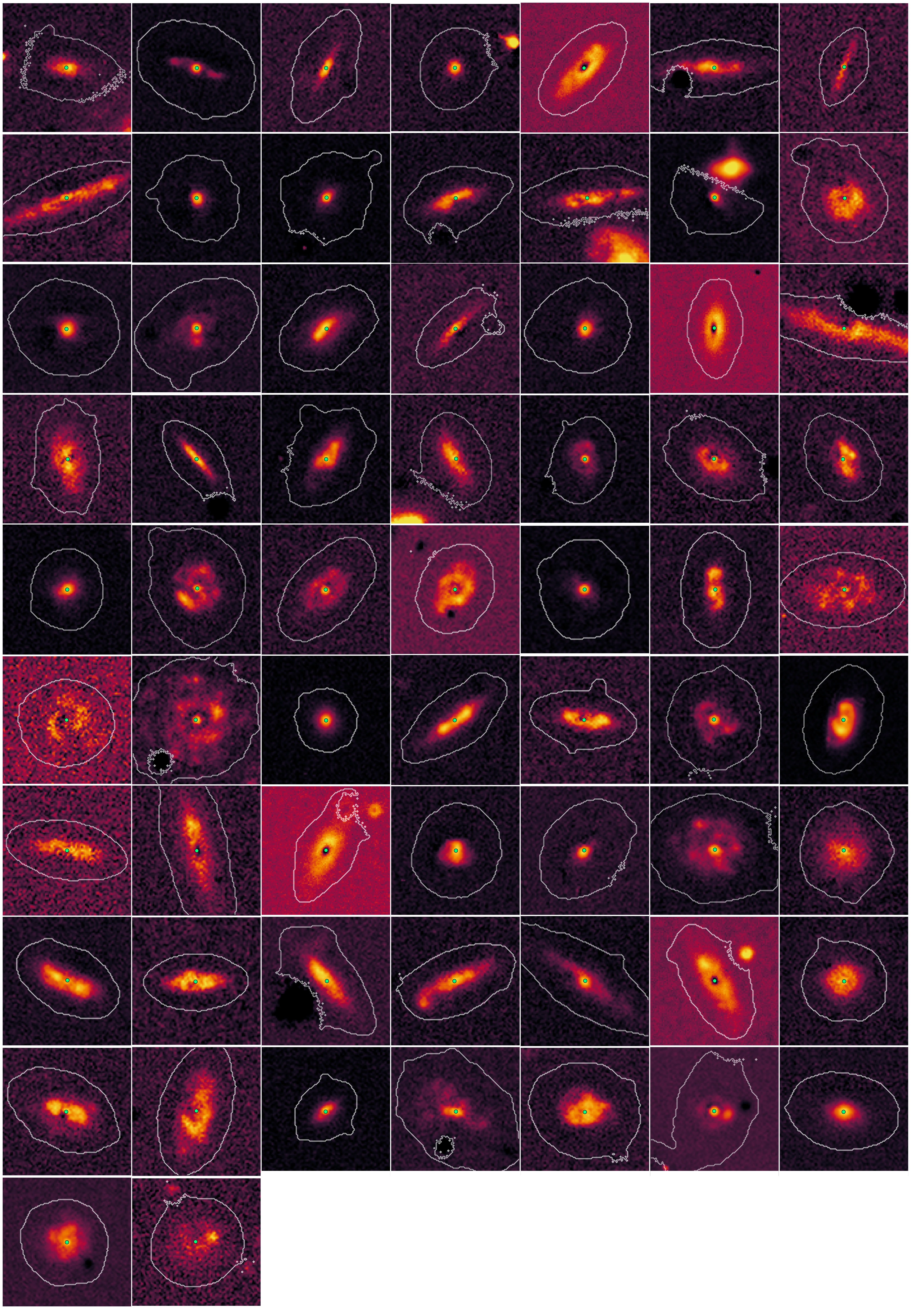}
\caption{Same as \autoref{fig:lowmasshighssfr} but for the galaxies in the highest-mass bin ($\log (M_\star/M_\odot) >10.1)$ with the lowest SSFR (bottom SSFR quartile of high-mass sources). The H$\alpha$ emission is generally more symmetric than that of the the sources shown in \autoref{fig:highmasshighssfr}. \label{fig:highmasslowssfr} }
\end{center}
\end{figure*}

To contextualize both the sample at hand and the nature of our mapping technique, we provide here a gallery of H$\alpha$ maps for various subsets of the dataset – complete sets of the selected galaxies are shown in Figure~\ref{fig:highlow_mass_ssfr}. In \autoref{fig:lowmasshighssfr} we show a selection of the lowest stellar mass ($\log (M_\star/M_\odot)<9.3$) and highest-SSFR galaxies. Conversely, in \autoref{fig:lowmasslowssfr} we show galaxies with low stellar masses and low SSFRs. We repeat this exercise for the high stellar mass galaxies ($\log (M_\star/M_\odot)>10.1$) with both high SSFRss (\autoref{fig:highmasshighssfr}) and low SSFRs (\autoref{fig:highmasslowssfr}).

\section{The Viability of a Centrally Concentrated Burst of Star Formation in a Low-mass Galaxy in Vertical Equilibrium}\label{s:appendix:prfm}
In Section~\ref{subsec:ha_conc} of the main text, we argue that some external nonequilibrium process drives the observed central starbursts that we observe in our low-mass galaxy sample. Here we provide a simple exercise to make the case against a purely secular or stochastic interpretation for these starbursts.

In the PRFM framework of star formation \citep{Ostriker2022}, a disk in a quasi-steady state should display a rough balance between the interstellar medium weight $\mathcal{W}$ (often approximated by the dynamical equilibrium pressure $\mathcal{W} \approx P_\textrm{DE} = \frac{1}{2}\pi G \Sigma_g^2 + \Sigma_g \sqrt{2 G \rho_{sd}}\sigma_{\rm eff}$, where $\Sigma_g$ is the total gas surface density, $\rho_{sd}$ is the combined stellar and dark matter density, and $\sigma_{\rm eff}$ is the effective velocity dispersion – assumed to be 10\,km\,s$^{-1}$ approximately the warm gas sound speed) and the pressure difference across the galaxy midplane. This framework has been shown to hold for non-starbursting low-mass galaxies wherein the ratio between weight and SFR surface density is characterized by $\Upsilon_{\rm tot} = {P_\textrm{DE}}/{\Sigma_{\rm SFR}} \sim 1000 \text{km s}^{-1}$ \citep{Kado-Fong2022}. Here we will consider whether the gaseous structure implied by a starbursting Merian dwarf galaxy could reasonably be attained in the vertical equilibrium case.

For an $M_\star = 10^9 M_\odot$ galaxy, let us adopt a stellar effective radius of $R_e = 2$~kpc with a central starburst that is 600\,pc in extent (i.e. $R_{\rm H\alpha}=0.6$\,kpc) and an SSFR~$=5\times 10^{-10}$~yr$^{-1}$. We follow the procedure outlined in \cite{Kado-Fong2022} to compute $\rho_{sd}$, assuming in particular that the ratio between the scale height and scale length is 0.3 (which itself follows from the results of \citet{Kado-Fong2020b}) and that the ratio between the stellar density and $\rho_\textrm{sd}$ is $\rho_\star/\rho_\textrm{sd}=0.46$. This allows us to solve for the required total gas surface density $\Sigma_g$ required to explain the observed average SFR surface density ($\Sigma_{\rm SFR} = {\rm SFR}/[\pi R_{\rm H\alpha}^2]$) given the observed average stellar mass surface density ($\Sigma_\star = M_\star / [\pi R_e^2]$). 

We estimate a required gas surface density in excess of 230~$M_\odot$\,pc$^{-2}$. However, from the results of \cite{Kado-Fong2022a} and \citet{Bradford2015} we expect the average atomic gas surface density of a typical gaseous disk to be to be around $\Sigma_{\rm HI}\sim 60$~$M_\odot$~pc$^{-2}$ given the stellar mass of the system. This would imply a molecular gas fraction exceeding $\Sigma_{\rm H_2}/\Sigma_g\approx0.75$. Although this high molecular gas fraction is possible in the centers of dwarf galaxies, the average molecular gas fraction in low-mass galaxies is expected to be lower by a factor of several \citep[see, e.g.][]{delosreyes2019,Kado-Fong2022}. 

From this simple exercise, we find that in the quasi-steady assumption a low-mass galaxy would require notably elevated molecular gas surface densities to sustain the highest specific star formation rates seen in our low-mass sample. It is thus reasonable to conclude that external processes such as mergers and and cold accretion play a significant role in powering the starbursts in our low-mass sample. Furthermore, although our toy model does not strictly hold for systems out of vertical equilibrium (e.g. post-merger systems), we note that this exercise also implies that, due to the diffuse and low-density structure of the low-mass systems, vigorous starbursts may only be viable at the galaxy center, where the molecular gas fraction is thought to be significantly elevated. Indeed, estimates using the molecular or extended Kennicutt-Schmidt relations \citep[following][]{delosreyes2019} imply even larger ($>0.9$) molecular gas fractions. These back-of-the-envelope exercises are consistent with our finding that high-SSFR, low-mass galaxies host significantly more concentrated star formation as compared to their high-SSFR, high-mass counterparts.

\begin{deluxetable*}{cccccccc}
\tablecaption{Full Spectroscopic Sample\label{tab:specsample}}
\tablewidth{0pt}
\tablehead{
\colhead{Merian ID} &\colhead{RA}	&\colhead{Dec}
&\colhead{$z_{spec}$} &\colhead{$M_\star$} &\colhead{$g-r$}&\colhead{$L_{r}$}& \colhead{Spec Source}\\[0cm]
\colhead{}&\colhead{(J2000)}	&\colhead{(J2000)}&\colhead{}&\colhead{$\log(M_\odot)$} &\colhead{mag}&\colhead{$\log(L_\odot)$}&\colhead{} \\[-0.1cm]
\colhead{(1)}&\colhead{(2)}&\colhead{(3)}&\colhead{(4)}&\colhead{(5)}&\colhead{(6)}&\colhead{(7)}&\colhead{(8)}
}
\startdata
2950279968392759821 &	 02:13:26.664 &	 -06:16:37.84 &	 0.0866 &	 9.53 &	0.49 &	 9.44 &	 GAMA \\
2950337142997422666 &	 02:10:41.515 &	 -06:05:18.68 &	 0.0870 &	 9.37 &	0.42 &	 9.39 &	 GAMA \\
2950354735183450022 &	 02:14:20.034 &	 -06:02:02.36 &	 0.0991 &	 9.28 &	0.40 &	 9.33 &	 GAMA \\
2950363531276481619 &	 02:12:52.955 &	 -05:57:21.25 &	 0.0904 &	 9.76 &	0.44 &	 9.75 &	 GAMA \\
2950605423834567700 &	 02:17:57.772 &	 -06:26:24.91 &	 0.0926 &	 9.62 &	0.32 &	 9.78 &	 GAMA \\
2950645006253186410 &	 02:18:03.492 &	 -06:24:22.62 &	 0.0921 &	 9.72 &	0.34 &	 9.85 &	 GAMA \\
2950671394532257761 &	 02:20:03.565 &	 -06:09:30.78 &	 0.0834 &	 9.51 &	0.67 &	 9.18 &	 GAMA \\
2950706578904338273 &	 02:14:31.173 &	 -06:08:12.87 &	 0.0930 &	 9.23 &	0.80 &	 8.74 &	 GAMA \\
2950710976950853296 &	 02:20:13.860 &	 -06:00:27.26 &	 0.0926 &	 9.03 &	0.37 &	 9.12 &	 GAMA \\
2950719773043882287 &	 02:19:00.808 &	 -05:57:50.89 &	 0.0856 &	 9.82 &	0.57 &	 9.63 &	 GAMA \\
\enddata
\tablenotetext{}{\textbf{Note.} The full spectroscopic catalog as described in Section~\ref{sec:data} including \specsamplesizeinbandlowmassallphot\ sources from the Merian photometric catalog with available spectroscopic redshifts.  The columns are: (1) unique identification number assigned to every Merian photometric source; (2) right ascension; (3) declination; (4) spectroscopic redshift; (5) stellar mass as calculated in Section~\ref{sec:data}; (6) $g-r$ color; (7) $r$-band luminosity; (8) source of spectrum.   (This table is available in its entirety in machine-readable form.)}
\end{deluxetable*}
\end{document}